\documentclass[11pt,a4paper]{article}
\usepackage{jheppub}

\pdfoutput=1
\usepackage{dcolumn}
\usepackage{bm}
\usepackage{here}
\usepackage{cancel} 
\usepackage{amsmath}
\usepackage{amssymb}  
\usepackage{slashed} 
\usepackage{booktabs}
\usepackage{tikz}
\usepackage[compat=1.1.0]{tikz-feynman}
\usepackage{tabularx}
\usepackage{amsmath,amssymb}
\usepackage{appendix}
\usepackage{graphicx}
\usepackage{color}
\usepackage{xcolor}
\usepackage{cancel}
\usepackage{subfigure}
\usepackage{multirow}
\usepackage{amssymb,changes}
\usepackage{hyperref}

\newcommand{\Rmnum}[1]{\expandafter\@slowromancap\romannumeral #1@}
\makeatother

\graphicspath{
	{./figure/}
}

\title{Dark Matter Production from Bubble Collisions during a First-Order Phase Transition at the End of Inflation}

\author{Zihong Cheng, }
\emailAdd{chengzh83@mail2.sysu.edu.cn}

\author{Fa Peng Huang\footnote{Corresponding author.}}
\emailAdd{huangfp8@sysu.edu.cn}

\affiliation{MOE Key Laboratory of TianQin Mission, TianQin Research Center for Gravitational Physics \& School of Physics and Astronomy, Frontiers Science Center for TianQin, Gravitational Wave Research Center of CNSA, Sun Yat-sen University (Zhuhai Campus), Zhuhai 519082, China}

\abstract{
We study whether a first-order phase transition at the end of inflation can generate the observed dark matter abundance through bubble collisions. The transition occurs in a spectator scalar sector with an inflaton-dependent effective potential, so that the nucleation rate grows during inflation and becomes significant only near its end. We identify the region of parameter space in which vacuum decay is dominated by the Coleman--De~Luccia channel, the Hawking--Moss transition remains subdominant, and the nucleated bubbles admit a consistent physical interpretation in an inflating background. Requiring also that the phase transition completes successfully, we then analyze particle production from bubble collisions. In the viable regime, elastic self-scatterings of the spectator particles can efficiently redistribute their momenta, while their decay into dark matter provides the dominant channel for transferring the spectator population to the dark sector. Other competing number-changing or sink processes remain inefficient compared with the Hubble expansion. The final relic abundance receives contributions from both direct production in bubble collisions and the subsequent decay of spectator field particles. We find that the observed dark matter abundance can be accommodated, within the order-of-magnitude accuracy of the collision-production treatment, in a restricted region of parameter space.
}


\begin{document}

\maketitle

\section{Introduction}

The particle nature and cosmological origin of dark matter (DM) remain among the central open questions in particle physics and cosmology \cite{Cirelli:2024ssz,Bozorgnia:2024pwk,Lin:2019uvt,Bertone:2016nfn,Boveia:2022adi}. 
While thermal freeze-out \cite{Gondolo:1990dk,Arcadi:2017kky,Roszkowski:2017nbc} and freeze-in \cite{McDonald:2001vt,Hall:2009bx,Bernal:2017kxu,Goudelis:2018xqi} provide the standard benchmarks for generating the DM relic abundance, the increasingly extensive experimental searches, together with the lack of positive signals in direct detection, indirect detection, and collider experiments, have motivated sustained interest in alternatives to the simplest thermal weakly interacting massive particle (WIMP) paradigm. 
In this context, cosmological first-order phase transitions (FOPTs) offer a particularly rich setting for DM genesis. 
FOPT can generate the DM abundance or substantially modify its final value through several distinct channels.
Ultra-relativistic bubble walls provide a non-thermal mechanism for producing particles far heavier than the phase transition scale, whereas bubble--plasma collisions furnish an additional efficient mechanism for populating secluded dark sectors \cite{Baldes:2023cih,Azatov:2021ifm,Azatov:2024crd}.
In addition, FOPTs may produce asymmetric DM or macroscopic non-thermal relics, such as (gauged) Q-ball and Fermi-ball DM \cite{Krylov:2013qe,Huang:2017kzu,Hall:2019rld,Hong:2020est,Jiang:2024zrb,Jiang:2025xln,Jiang:2023qbm}.
Distinct from these scenarios, bubble wall filtering can also set the relic abundance of particle DM: if the DM mass changes across the wall, only sufficiently energetic particles can enter the broken phase, while the rest are reflected and depleted, realizing the filtered DM mechanism \cite{Baker:2019ndr,Chway:2019kft,Jiang:2023nkj}. 
Against this broader backdrop, bubble collisions constitute another especially compelling channel, as they can directly convert the energy stored in the expanding bubble walls into DM particles and thereby provide a distinct non-thermal production mechanism \cite{Watkins:1991zt,Falkowski:2012fb,Mansour:2023fwj,Giudice:2024tcp,Cataldi:2025nac,Freese:2023fcr}. 
Among the possible realizations of this mechanism, a particularly intriguing one is that the FOPT takes place already during inflation rather than after reheating. 
In such a setup, an inflaton-induced FOPT in a spectator sector can produce superheavy DM far from equilibrium, as pointed out in Ref.~\cite{An:2022toi}. 
Motivated by this possibility, in this work we study DM production from a FOPT during inflation in greater detail. 

A FOPT proceeds through the nucleation of true vacuum bubbles inside a metastable false vacuum. 
In the scenario considered here, however, the transition occurs during inflation, so the bubbles expand in an approximately empty vacuum rather than in a thermal plasma. 
As a consequence, plasma friction is absent and the bubble walls can continue to accelerate until they enter an ultrarelativistic runaway regime before collision \cite{An:2022toi,Cataldi:2025nac,Espinosa:2010hh,Konstandin:2010dm,Bodeker:2009qy}. 
The vacuum energy stored in these highly boosted walls is then released in bubble collisions, efficiently exciting fields coupled to the spectator sector \cite{Falkowski:2012fb,Mansour:2023fwj,Giudice:2024tcp,Cataldi:2025nac}. 
If the spectator field is coupled to a stable dark sector particle, these non-thermal processes can provide an efficient source of DM. 
Our focus is therefore on the regime of an inflationary vacuum with runaway bubbles, in which DM production is governed by vacuum bubble dynamics rather than by thermal scattering after reheating \cite{An:2022toi}.

This focus immediately raises a first central issue: in an inflationary background, the vacuum decay problem is qualitatively different from its familiar flat-spacetime counterpart. In Minkowski space, false vacuum decay is described by the semiclassical bounce formalism of Refs.~\cite{Coleman:1977py,Callan:1977pt}, with explicit quartic potential solutions given in Ref.~\cite{Adams:1993zs}. Once gravitational effects are included, however, the relevant tunneling process must be analyzed in an approximately de~Sitter background, where the structure of Euclidean saddle points can differ substantially from that in flat spacetime \cite{Coleman:1980aw,Sasaki:1994yj}. Depending on the shape of the barrier and its relation to the Hubble scale, a Coleman--De~Luccia (CDL) bounce may dominate, may coexist with the Hawking--Moss solution \cite{Weinberg:2006pc,Hawking:1981fz}, or may even cease to exist \cite{Balek:2003uu,Joti:2017fwe,Vicentini:2022pra}. This distinction is crucial for the present work, because the DM production mechanism of interest requires a genuine localized bubble nucleation event, followed by bubble expansion and subsequent collisions, rather than a homogeneous transition over the barrier. At the same time, identifying the physically relevant tunneling saddle is subtle in de~Sitter space, since some Euclidean solutions possess additional negative modes and therefore cannot be straightforwardly interpreted as the tunneling saddle \cite{Lee:2014uza,Jinno:2020zzs,Battarra:2012vu,Bramberger:2019mkv}.  We therefore analyze vacuum decay directly in de~Sitter space, rather than relying on the flat-space approximation, and identify the relevant saddle in the parameter region of interest. In particular, we restrict attention to the parameter regime in which localized bubble nucleation is the relevant semiclassical description of the transition.

A second central issue is whether the phase transition can occur during inflation without either disrupting the inflationary background too early or failing to complete near its end. 
If bubble nucleation becomes efficient too early, large bubbles may form while inflation is still ongoing and generate unacceptable inhomogeneities. 
Conversely, if the nucleation rate rises too slowly, then even if bubbles are eventually produced, accelerated expansion can separate them faster than they can grow and collide with one another, thereby preventing efficient percolation and completion of the phase transition \cite{Turner:1992tz,Ellis:2018mja}. 
A viable scenario therefore requires the nucleation rate to remain sufficiently suppressed throughout most of the inflationary era, and then to increase rapidly near the end of inflation so that bubbles nucleate abundantly and collide efficiently within a Hubble time. 
In practice, this imposes nontrivial constraints on the time dependence of the tunneling exponent and on the dimensionless parameter \(\beta/H\), which measures the timescale over which the nucleation rate rises relative to the Hubble expansion.

A third essential aspect of the problem is the connection between particle production at the phase transition and the present-day DM abundance.
It is not sufficient to show only that a FOPT can take place during inflation. 
One must also track the subsequent evolution of the particles produced in bubble collisions and in the decay or other dynamical processes of the spectator field, including their dilution during the remaining inflationary stage and their later cosmological redshifting. 
A viable scenario therefore requires a careful relation between the production yield at the time of the phase transition, the number of remaining $e$-folds after production, and the relic density observed today \cite{An:2022toi}.

For concreteness, throughout this work we assume instantaneous reheating followed by the standard radiation-dominated era, so that the inflaton energy density is promptly converted into radiation after inflation ends. 
Within this framework, we present a unified analysis of DM production from a FOPT during inflation. 
We specify a setup in which a DM sector is coupled to the spectator field, study vacuum decay in de~Sitter space with particular emphasis on the existence and physical relevance of the CDL and Hawking--Moss solutions, review the conditions for a cosmologically viable transition during inflation, and compute the resulting DM abundance from both bubble collisions and the subsequent decay of the spectator field. 
We then identify the parameter region consistent with the observed relic abundance and evaluate the associated gravitational wave (GW) signal.

The rest of this paper is organized as follows. 
In Sec.~\ref{sec:model}, we introduce the model, establish our notation, and specify the inflationary setup. 
In Sec.~\ref{sec:vacdecay}, we review and extend the analysis of vacuum decay in de~Sitter space. 
In Sec.~\ref{sec:DMproduction}, we compute the DM abundance produced by bubble collisions and by the decay of the spectator field. 
In Sec.~\ref{sec:GW}, we discuss the accompanying GW signal. 
Finally, we summarize our conclusions in Sec.~\ref{sec:conclusion}. 
Additional details are collected in the appendices.

\section{Model setup}
\label{sec:model}

Before analyzing vacuum decay and particle production, we first specify the model setup and the approximations adopted throughout this work. 
In particular, we introduce the particle content and interactions, characterize the spectator sector potential and its vacuum structure, and summarize the inflationary background relevant for the phase transition. 
These ingredients will form the basis of the de~Sitter vacuum decay analysis in Sec.~\ref{sec:vacdecay} and the computation of DM production induced by bubble collisions in Sec.~\ref{sec:DMproduction}.

\subsection{Particle content and interactions}

We consider an inflationary background driven by the inflaton field $\phi$, together with a real scalar spectator field $\sigma$ that undergoes a FOPT during inflation. 
We do not specify the microscopic realization of inflation and instead treat the Hubble parameter $H$ as an external input.
Accordingly, $\phi$ is regarded as a homogeneous slow-roll background. 
DM is taken to be a fermion $\psi$, coupled to the spectator field through a Yukawa interaction.

The action describing gravity and the two-scalar system is given by
\begin{equation}
S
=
\int d^4x \sqrt{-g}\,
\left[
\frac{M_{\rm Pl}^2}{2}\,R
+\frac{1}{2}\,g^{\mu\nu}\partial_\mu \phi\,\partial_\nu \phi
- V_{\rm inf}(\phi)
+\frac{1}{2}\,g^{\mu\nu}\partial_\mu \sigma\,\partial_\nu \sigma
- V(\phi,\sigma)
\right] ,
\label{eq:S-full}
\end{equation}
Throughout this work, we adopt the metric signature \(g_{\mu\nu}=\mathrm{diag}(+,-,-,-)\),  where \(V_{\rm inf}(\phi)\) denotes the inflaton potential, and \(V(\phi,\sigma)\) is the spectator sector potential.

We take the effective potential for the spectator field to be~\cite{An:2020fff,An:2022toi}
\begin{equation}
V(\phi,\sigma)
=
\frac{1}{2}\,m^2(\phi)\,\sigma^2
+\frac{\lambda\epsilon}{3}\,\sigma^3
+\frac{\kappa}{4}\,\sigma^4 ,
\label{eq:Vphisigma}
\end{equation}
with
\begin{equation}
m^2(\phi)\equiv \mu^2-c_{\rm inf}^2\phi^2 .
\label{eq:mueff}
\end{equation}
We fix $\lambda=-1$ and take $\epsilon>0$, $\mu>0$, $c_{\rm inf}>0$, and $\kappa>0$ throughout. Here $\epsilon$ and $\mu$ carry mass dimension one, while $c_{\rm inf}$ and $\kappa$ are dimensionless.
For a fixed background value of $\phi$, and with $\kappa>0$ to ensure stability at large field values, the potential can then develop the characteristic barrier structure required for a FOPT. 
More specifically, for suitable values of $m^2(\phi)$ and $\epsilon$, it admits a metastable false vacuum separated from the true vacuum by a finite potential barrier. 
The role of the inflaton background is to induce a slow time dependence in $m^2(\phi)$, thereby making the tunneling rate evolve during inflation.

We further take the DM particle to be a fermion $\psi$, with dark sector Lagrangian
\begin{equation}
\mathcal{L}_{\rm DM}
=
\bar\psi \left(i\gamma^\mu \partial_\mu - m_\psi - y\,\sigma\right)\psi .
\label{eq:LDM}
\end{equation}
Equivalently, the interaction between the spectator field and DM is described by the Yukawa term
\begin{equation}
\mathcal{L}_{\rm int}
=
-y\,\sigma\,\bar\psi\psi .
\label{eq:Lint}
\end{equation}
 This coupling opens multiple channels for DM production during the phase transition, including direct production from bubble collisions and indirect production through the subsequent decay of spectator field particles. As will be discussed below, a necessary condition for the phase transition to complete during inflation is \(\beta/H > 10\). In the parameter region actually used for our numerical analysis, however, we find the stronger hierarchy \(\beta/H \gtrsim 10^2\). This means that the characteristic time scale of bubble collisions, \(\Delta t_{\rm coll} \sim \beta^{-1}\), is much shorter than the Hubble time, so that \(H \Delta t_{\rm coll} \sim H/\beta \lesssim 10^{-2} \ll 1\). Likewise, the typical physical size of the collision region is \(R_{\rm coll} \sim v_w \beta^{-1} \lesssim \beta^{-1}\),  implying \(H R_{\rm coll} \lesssim H/\beta \lesssim 10^{-2} \ll 1\). Therefore, in the parameter space actually considered in this work, both the time and length scales associated with the microscopic bubble collision process are much smaller than the Hubble scale \(H^{-1}\). We thus compute the DM production induced by bubble collisions in the flat-spacetime approximation. This approximation is applied only to the local collision dynamics and the associated particle production process, while the subsequent cosmological evolution of the produced DM is still treated in the expanding background.

\subsection{Potential shape and vacuum structure}

For both the analytical treatment of vacuum decay and the numerical scan, it is useful to characterize the spectator potential in terms of the locations of its extrema and the corresponding vacuum energy difference. 
In the following, we therefore use the potential defined in Eqs.~\eqref{eq:Vphisigma} and~\eqref{eq:mueff}, fixing the background inflaton value $\phi$ and treating it as an external parameter. 
With $\lambda=-1$, the cubic term provides the negative contribution needed to generate a barrier, while $\kappa>0$ ensures stability at large field values. 
For suitable values of $m^2(\phi)$ and $\epsilon$, the potential then admits the barrier structure required for a FOPT, with a metastable false vacuum separated from the true vacuum by a finite potential barrier.

The stationary points are determined by
\begin{equation}
\frac{\partial V}{\partial \sigma}
=
\sigma\left[m^2(\phi)-\epsilon\,\sigma+\kappa\,\sigma^2\right]
=0 .
\end{equation}
Besides the extremum at the origin,
\begin{equation}
\sigma_0=0 ,
\end{equation}
the nontrivial stationary points are
\begin{equation}
\sigma_\pm
=
\frac{\epsilon\pm\sqrt{\Delta(\phi)}}{2\kappa},
\qquad
\Delta(\phi)\equiv \epsilon^2-4\kappa m^2(\phi) .
\end{equation}
A barrier exists when the nontrivial extrema are real and the origin remains a local minimum, namely in the range
\begin{equation}
0<m^2(\phi)<\frac{\epsilon^2}{4\kappa}.
\end{equation}
In this regime one has $0<\sigma_-<\sigma_+$, where $\sigma_-$ is the barrier top and $\sigma_+$ is the nonzero local minimum. 
It is therefore convenient to identify
\begin{equation}
\sigma_{\rm top}\equiv \sigma_-,
\qquad
\sigma_{\rm nz}\equiv \sigma_+ .
\end{equation}
In the parameter region where \(m^2(\phi)<0\), the effective mass squared becomes negative, and the dynamics are correspondingly modified, the origin is tachyonic rather than metastable, while for \(m^2(\phi)>\epsilon^2/(4\kappa)\), the nonzero stationary points disappear and the barrier no longer exists.

The potential values at the relevant extrema are
\begin{align}
V(0)&=0,\\
V(\sigma_{\rm top})
&=
\frac{(\epsilon-\sqrt{\Delta})^3(\epsilon+3\sqrt{\Delta})}{192\,\kappa^3},\\
V(\sigma_{\rm nz})
&=
\frac{(\epsilon+\sqrt{\Delta})^3(\epsilon-3\sqrt{\Delta})}{192\,\kappa^3}.
\end{align}
The two local minima become degenerate when \(V(\sigma_{\rm nz})=V(0)=0\), which gives \(m^2(\phi)=2\epsilon^2/(9\kappa)\). Within the two-minimum regime, for \(0<m^2(\phi)<2\epsilon^2/(9\kappa)\), the nonzero minimum is deeper, \(V(\sigma_{\rm nz})<0\), so that \(\sigma=\sigma_{\rm nz}\) is the true vacuum and \(\sigma=0\) is the false vacuum. By contrast, in the range \(2\epsilon^2/(9\kappa)<m^2(\phi)<\epsilon^2/(4\kappa)\), one has \(V(\sigma_{\rm nz})>0\), and the roles of the two minima are exchanged: the origin becomes the true vacuum, while the nonzero minimum becomes metastable.

Since our main interest is the late-time tunneling process, we focus on this metastable regime and identify
\begin{equation}
\sigma_{\rm tv}=0,
\qquad
\sigma_{\rm fv}=\sigma_+
=\frac{\epsilon+\sqrt{\Delta}}{2\kappa},
\qquad
\sigma_{\rm top}=\sigma_-
=\frac{\epsilon-\sqrt{\Delta}}{2\kappa}.
\end{equation}
The corresponding vacuum-energy difference is
\begin{equation}
\Delta V
\equiv
V(\sigma_{\rm fv})-V(\sigma_{\rm tv})
=
V(\sigma_+)-V(0)
=
\frac{(\epsilon+\sqrt{\Delta})^3(\epsilon-3\sqrt{\Delta})}{192\,\kappa^3}.
\end{equation}
In the metastable regime considered here, this quantity is positive, as required.

The local curvatures at the relevant extrema are
\begin{align}
V''(0)&=m^2(\phi),\\
V''(\sigma_\pm)
&=
\frac{\Delta \pm \epsilon\sqrt{\Delta}}{2\kappa}.
\end{align}
Hence,
\begin{equation}
V''(\sigma_{\rm fv})=V''(\sigma_+)
=\frac{\Delta+\epsilon\sqrt{\Delta}}{2\kappa}>0,
\qquad
V''(\sigma_{\rm top})=V''(\sigma_-)
=\frac{\Delta-\epsilon\sqrt{\Delta}}{2\kappa}<0,
\end{equation}
which confirms that $\sigma_{\rm fv}$ is a local minimum and $\sigma_{\rm top}$ is the barrier top. 
The set of quantities
\begin{equation}
\{\sigma_{\rm fv},\,\sigma_{\rm tv},\,\sigma_{\rm top},\,\Delta V\}
\end{equation}
then provides a convenient characterization of the potential for both the analytical study of vacuum decay and the numerical evaluation of the tunneling dynamics.

\subsection{Inflationary background and approximations}

The background geometry is well approximated by de~Sitter space,
\begin{equation}
a(t)=e^{Ht},
\end{equation}
with the Hubble parameter \(H\) remaining approximately constant throughout the inflationary epoch. 
Rather than specifying a concrete microscopic model of inflation, we treat the inflaton as a homogeneous slow-roll background and use an effective parameterization of its evolution.

Starting from the slow-roll expression for the number of $e$-folds,
\begin{equation}
N(\phi)= \int_{\phi_{\rm end}}^{\phi}\frac{d\phi}{M_{\rm Pl}\sqrt{2\epsilon_V(\phi)}} ,
\end{equation}
and approximating the evolution over the relevant interval by an effective average slow-roll parameter $\bar\epsilon_V$, we obtain
\begin{equation}
\Delta\phi \simeq N\,M_{\rm Pl}\sqrt{2\bar\epsilon_V},
\qquad
\phi(N)\simeq \phi_{\rm end}+N\,M_{\rm Pl}\sqrt{2\bar\epsilon_V},
\label{eq:phiN-approx}
\end{equation}
where $M_{\rm Pl}$ is the reduced Planck mass, defined by
$M_{\rm Pl}^{-2}\equiv 8\pi G.$

In the numerical analysis, we take $\bar\epsilon_V=0.001$ and $\phi_{\rm end}=0$ as a benchmark parameterization of the slow-roll evolution, with $\bar\epsilon_V = 0.001$ chosen near the current observational upper bound~\cite{Balkenhol:2025wms} . 
This allows us to rewrite the spectator potential $V(\phi,\sigma)$ as an effective $N$-dependent potential and to follow the evolution of the false vacuum, the true vacuum, and the barrier as inflation proceeds, as illustrated in Fig.~\ref{figvaccum}.

Within this setup, we further adopt the following approximations throughout this work. 
We neglect the backreaction of the spectator sector on the inflationary background, so that the inflaton enters the analysis only through the time dependence of $m^2(\phi)$ in Eq.~\eqref{eq:mueff}. 
For the post-inflationary evolution, we assume instantaneous reheating followed by a radiation-dominated era, such that the inflaton energy density is promptly converted into radiation at the end of inflation.

\section{Vacuum decay and bubble nucleation during inflation}
\label{sec:vacdecay}

Before computing DM production from bubble collisions, we must first establish that the underlying phase transition can proceed through physically meaningful localized bubble nucleation during inflation. 
This requires more than the existence of a metastable potential. In an inflationary background, one must identify the relevant tunneling saddle in de~Sitter space, determine whether a localized CDL bounce exists and remains dominant over the Hawking--Moss channel, and ensure that the resulting nucleation history is compatible with the cosmological requirements of inflation. 
This point is particularly important because a flat-space bounce approximation can be misleading in part of the parameter space: it may overestimate the region that admits viable bubble nucleation, since gravitational effects can qualitatively modify the decay structure, including the possible disappearance of the CDL solution or the relevance of the Hawking--Moss transition. 
Moreover, even when a CDL bounce does exist, its profile and tunneling exponent can differ appreciably from those obtained in the flat-space approximation. 
For this reason, we begin with a brief review of false vacuum decay in flat space only as a reference point, and then develop the corresponding analysis in de~Sitter space. 
We next discuss the conditions for the existence and disappearance of the CDL solution, the constraints on the nucleation rate at different inflationary epochs, and the practical framework used in our numerical evaluation of the decay rate.

\subsection{Vacuum decay in flat space}

Before turning to vacuum decay in de~Sitter space, we briefly review the standard flat-space formalism of false vacuum decay~\cite{Coleman:1977py,Callan:1977pt,Coleman:1985rnk,Lee:2014uza}. 
Throughout this section, without loss of generality, we denote the field undergoing the phase transition by \(\varphi\).

In flat space, the decay of a metastable vacuum can be formulated in terms of the Euclidean transition amplitude from the false vacuum back to itself,
\begin{equation}
Z(T)\equiv
\langle \varphi_{\rm fv} | e^{-HT} | \varphi_{\rm fv} \rangle
=
\int [d\varphi]\, e^{-S_E[\varphi]},
\end{equation}
where \(T\) denotes the Euclidean time interval rather than the thermodynamic temperature, and
\begin{equation}
S_E[\varphi]
=
\int d\tau\, d^3x\,
\left[
\frac12 \left(\partial_\tau \varphi\right)^2
+\frac12 (\nabla \varphi)^2
+U(\varphi)
\right]
\end{equation}
is the Euclidean action. 
In the large-\(T\) limit, \(Z(T)\) is dominated by the lowest energy state with nonzero overlap with \(|\varphi_{\rm fv}\rangle\). 
Identifying this state with the metastable false vacuum, one is led to define its energy as
\begin{equation}
E_{\rm fv}
=
-\lim_{T\to\infty}\frac{1}{T}\ln Z(T).
\label{eq:false_vac_energy_def}
\end{equation}
If \(E_{\rm fv}\) develops an imaginary part, the metastable state is unstable, and the decay rate is determined by that imaginary part.

Semiclassically, the path integral is evaluated by expanding around the nontrivial \(O(4)\)-symmetric Euclidean bounce solution \(\varphi_b\), which describes the nucleation of a critical bubble. 
The corresponding tunneling exponent is
\begin{equation}
B \equiv S_E[\varphi_b]-S_E[\varphi_{\rm fv}].
\end{equation}
A crucial point is that the bounce is not a minimum of the Euclidean action, but a saddle point. 
Its fluctuation operator contains four zero modes associated with translations and, more importantly, a single negative mode. The latter contributes a factor of \(i\) upon Gaussian integration.
It is precisely this factor that gives rise to the imaginary part of \(E_{\rm fv}\), thereby allowing the bounce to be interpreted as the semiclassical saddle governing false vacuum decay.

After summing over dilute multi-bounce configurations, one obtains
\begin{equation}
Z(T)=Z_0 \exp\!\left(i\,V T D e^{-B}\right),
\end{equation}
where \(V\) is the spatial volume and
\begin{equation}
D
=
\frac{J}{2}
\left|
\frac{\det{}' S_E''(\varphi_b)}{\det S_E''(\varphi_{\rm fv})}
\right|^{-1/2}.
\end{equation}
Here \(J\) is the Jacobian associated with the collective coordinates, and the prime indicates that the translational zero modes are omitted from the determinant. 
Substituting this result into Eq.~\eqref{eq:false_vac_energy_def}, one finds that the false vacuum energy becomes complex, with
\begin{equation}
\Gamma
=
-\frac{2}{V}\,\mathrm{Im}\,E_{\rm fv}
=
2D e^{-B}.
\end{equation}
Hence, in flat space, the existence of exactly one negative mode is not merely a mathematical property of the bounce solution; within the standard semiclassical treatment, it is the key condition that allows the bounce to be identified as the tunneling saddle responsible for false vacuum decay~\cite{Coleman:1987rm}.

\subsection{Vacuum decay in de Sitter space}

For false vacuum decay from a de~Sitter vacuum, gravitational effects are described by the CDL bounce on a compact Euclidean geometry~\cite{Coleman:1980aw,Tanaka:1992zw}. In what follows, we build upon the theoretical framework established in Ref.~\cite{Lee:2014uza}.
Assuming $O(4)$ symmetry, the Euclidean metric can be written as
\begin{equation}
ds^2=\mathcal{N}(\xi)\,d\xi^2+\varrho ^2(\xi)\,d\Omega_3^2,
\end{equation}
where \(\xi\) is the Euclidean radial coordinate and \(d\Omega_3^2\) is the metric on the unit three-sphere.
For a scalar field \(\varphi\) minimally coupled to gravity, the Euclidean action is
\begin{equation}
S_E
=
\int d^4x \sqrt{g}\,
\left[
-\frac{M_{\rm Pl}^2}{2}R
+\frac12 (\nabla\varphi)^2
+U(\varphi)
\right]
+S_{\rm bd},
\end{equation}
where \(U(\varphi)\) is the scalar potential and \(S_{\rm bd}\) denotes the Euclidean Gibbons--Hawking boundary term \cite{Lee:2014uza,Gibbons:1976ue}. 
For the \(O(4)\)-symmetric ansatz, the action reduces to
\begin{equation}
S_E
=
2\pi^2\int d\xi\,\sqrt{\mathcal N}\,
\left\{
-3M_{\rm Pl}^2
\left(
\frac{\varrho \dot\varrho ^{\,2}}{\mathcal N}+\varrho 
\right)
+\varrho ^3
\left[
\frac{\dot\varphi^{\,2}}{2\mathcal N}+U(\varphi)
\right]
\right\},
\end{equation}
where overdots denote derivatives with respect to \(\xi\).
Choosing the gauge \(\mathcal N=1\), one convenient set of background equations is
\begin{equation}
\dot\varrho ^{\,2}
=
1+\frac{\varrho ^2}{3M_{\rm Pl}^2}
\left(
\frac12\dot\varphi^{\,2}-U
\right),
\qquad
\ddot\varphi+3\frac{\dot\varrho }{\varrho }\dot\varphi
=
\frac{dU}{d\varphi}.
\end{equation}

Since our main interest is the negative mode structure rather than a precise evaluation of the tunneling exponent, we focus on quadratic fluctuations around the CDL bounce.
Using the gauge invariant fluctuation variable \(\chi\) introduced in Ref.~\cite{Lee:2014uza}, the quadratic action can be written as
\begin{equation}
S_E^{(2)}
=
2\pi^2\int d\xi\,L_E^{(2)},
\qquad
L_E^{(2)}
=
\frac{\varrho ^3}{2Q}\dot\chi^{\,2}
+
\frac{\varrho ^3}{2Q}g(\varrho ,\varphi)\chi^2,
\end{equation}
with
\begin{equation}
Q
\equiv
1-\frac{\varrho ^2U}{3M_{\rm Pl}^2}
=
\dot\varrho ^{\,2}
-\frac{\varrho ^2\dot\varphi^{\,2}}{6M_{\rm Pl}^2}.
\end{equation}
Here \(g(\varrho ,\varphi)\) denotes the coefficient of \(\chi^2\) in the gauge invariant quadratic Euclidean Lagrangian. In the present discussion, we do not need their explicit forms. Instead, the key quantity is \(Q\), whose sign plays a central role in diagnosing the negative mode structure. 
When \(Q<0\) in some region, the kinetic term changes sign there, and rapidly oscillating negative modes can appear in addition to the ordinary tunneling mode \cite{Lee:2014uza}.

Following Ref.~\cite{Lee:2014uza}, it is useful to distinguish the standard slowly varying negative mode from the rapidly oscillating negative modes associated with regions where \(Q<0\).
The former is connected to the familiar flat-space negative mode and underlies the usual semiclassical interpretation of the CDL bounce as a tunneling saddle.
The latter are specific to the gravitational problem and reflect the nonstandard negative mode structure that can arise in de~Sitter space.
Here we focus on the $O(4)$-symmetric scalar fluctuation sector, corresponding to the $\ell=0$ mode, where $\ell$ labels the angular momentum of the perturbation on the unit three-sphere. For the CDL bounce, this sector contains the standard tunneling negative mode as well as the rapidly oscillating negative modes associated with $Q<0$, whereas the $\ell=1$ sector does not contain independent gauge-invariant modes and the special negative kinetic term pathology is absent for $\ell\ge2$ \cite{Lee:2014uza}.

It is also useful to distinguish between different classes of de~Sitter bounces. 
Following Ref.~\cite{Lee:2014uza}, we define the characteristic barrier scale through 
\(U_{\rm top}-U(\varphi_{\rm fv}) \equiv \eta^4\), where \(U_{\rm top}\) is the potential at the top of the barrier, and the vacuum-energy difference as 
\(\Delta U \equiv U(\varphi_{\rm fv})-U(\varphi_{\rm tv})\), with \(\varphi_{\rm fv}\) and \(\varphi_{\rm tv}\) denoting the false and true vacuum field values, respectively. 
In the thin-wall approximation, the bubble radius at nucleation is given by
\begin{equation}
\frac{1}{\bar\varrho ^{\,2}}
=
H^2
+
\left(
\frac{\Delta U}{3\sigma_{\rm ten}} - \frac{\sigma_{\rm ten}}{4M_{\rm Pl}^2}
\right)^2,
\end{equation}
where \(H^2 = U(\varphi_{\rm fv})/(3M_{\rm Pl}^2)\) is the false vacuum Hubble parameter and \(\sigma_{\rm ten}\) is the bubble-wall tension. 
As \(\Delta U\) decreases, the bubble radius first increases up to \(\bar\varrho  = H^{-1}\) and then decreases again. 
Following Ref.~\cite{Lee:2014uza}, the first branch is termed type A, while the second is termed type B. 
More generally, a bounce is classified as type B if the maximum of \(\varrho \) lies within the wall region, and as type A otherwise.

Of particular importance for the present work is the regime of small type A bounces. 
According to Ref.~\cite{Lee:2014uza}, this corresponds to
\begin{equation}
\eta \ll M_{\rm Pl},
\qquad
\Delta U \ \text{not parametrically small},
\label{gravitycdlcondition}
\end{equation}
for which the nucleated bubble is much smaller than the de~Sitter radius, \(\bar\varrho  \ll H^{-1}\). 
In this regime, the bounce can be identified with a small bubble CDL solution in de~Sitter space. 
Crucially, this class of solutions retains the standard slowly varying negative mode, the bounce continues to admit the usual semiclassical interpretation as the saddle point describing vacuum tunneling, in close analogy with the flat-space case \cite{Lee:2014uza}.

As summarized in Table~I of Ref.~\cite{Lee:2014uza}, the negative mode content of de~Sitter bounces depends sensitively on the bounce class and on the sign of \(Q\). 
Small type-A bounces do not exhibit wall-localized oscillating modes, and the remaining oscillating modes are confined to an exponentially narrow region near \(\varrho _{\max}\). 
Large type-A bounces can additionally develop wall oscillating modes if \(Q\) becomes negative in the wall region. 
For type-B bounces, however, the standard slowly varying negative mode is absent, and the distinction between wall and \(\varrho _{\max}\)-localized oscillating modes ceases to be particularly meaningful.

We emphasize that the negative mode issue in gravitational vacuum decay is not yet fully understood at a fundamental level. 
As argued in Ref.~\cite{Jinno:2020zzs}, the ambiguity cannot be regarded as a purely technical artifact of variable choice, since different canonical formulations may lead to different negative mode criteria without a unique physical prescription. 
More broadly, this problem should be viewed in the context of the limitations of the semiclassical Euclidean treatment of gravity itself. 
In particular, the fact that the Euclidean gravitational action is unbounded from below~\cite{Gibbons:1978ac} already indicates that the gravitational path integral is conceptually subtle. 
It is therefore plausible that a clearer physical interpretation of vacuum decay with gravity will require frameworks beyond the standard Euclidean saddle point approach, including, for example, Wheeler--DeWitt-based analyses~\cite{DeAlwis:2019rxg,Jinno:2020zzs}.

In view of the discussion above, in the present work we follow the perspective of Ref.~\cite{Lee:2014uza} and restrict our analysis to the regime of small type-A bounces, as defined by Eq.~\eqref{gravitycdlcondition}. 
In this regime, the only negative mode that is physically relevant for our analysis is the standard slowly varying CDL mode, while the additional oscillating modes are confined to an exponentially narrow region near $\varrho _{\max}$ and are associated with trans-Planckian physics. 
Since our analysis is restricted to the semiclassical regime well below the Planck scale, we neglect such trans-Planckian effects and assume that no additional negative modes are relevant for the vacuum decay process. 
We therefore focus on the standard negative mode of the CDL bounce, namely the one corresponding to the familiar flat-space negative mode in the presence of gravity. 
Whether this standard negative mode is present or absent for a given solution will be determined later, where we will present the corresponding criterion explicitly.

\subsection{The Hawking--Moss instanton}

In flat spacetime, the homogeneous field configuration sitting at the top of the potential barrier does not constitute a finite action bounce. 
Its Euclidean action difference relative to the false vacuum is proportional to the infinite Euclidean four-volume, and thus it does not contribute to vacuum decay. 
In de~Sitter space, by contrast, the Euclidean background is a compact four-sphere with finite volume, so the corresponding homogeneous configuration has a finite action and can provide a relevant decay channel~\cite{Weinberg:2006pc,Hawking:1981fz,Sasaki:1994yj,Joti:2017fwe,Gregory:2020hia}.

The Hawking--Moss instanton is the corresponding homogeneous Euclidean solution, for which the scalar field remains constant over the entire Euclidean de~Sitter manifold and sits at the top of the potential barrier, \(\varphi=\varphi_{\mathrm{top}}\), satisfying
\begin{equation}
U'(\varphi_{\mathrm{top}})=0,
\qquad
U''(\varphi_{\mathrm{top}})<0.
\end{equation}
It is interpreted not as the nucleation of a localized bubble, but rather as a Hubble patch fluctuating to the top of the barrier~\cite{Weinberg:2006pc,Joti:2017fwe}.

For a minimally coupled scalar field, the Hawking--Moss tunneling exponent is given by the Euclidean action difference between the false vacuum and the homogeneous configuration at the top of the barrier,
\begin{equation}
B_{\rm HM}
=
24\pi^2 M_{\rm Pl}^4
\left[
\frac{1}{U(\varphi_{\rm fv})}
-
\frac{1}{U(\varphi_{\mathrm{top}})}
\right].
\end{equation}
Writing
\begin{equation}
U(\varphi)=U(\varphi_{\rm fv})+\delta U(\varphi),
\end{equation}
and evaluating the Hubble parameter at the false vacuum,
\begin{equation}
H^2=\frac{U(\varphi_{\rm fv})}{3M_{\rm Pl}^2},
\end{equation}
one obtains, for $\delta U(\varphi_{\mathrm{top}}) \ll U(\varphi_{\rm fv})$,
\begin{equation}
B_{\rm HM}
\simeq
\frac{8\pi^2}{3H^4}\,\delta U(\varphi_{\mathrm{top}}).
\end{equation}
This form makes explicit that the Hawking--Moss action scales as $H^{-4}$ and therefore becomes irrelevant in the flat-space limit $H\to 0$, while it can become important when the Hubble scale is sufficiently large or the barrier is sufficiently broad \cite{Sasaki:1994yj,Joti:2017fwe}.

For the scenario studied in this work, this distinction is crucial. 
Our DM production mechanism relies on the nucleation of localized true vacuum bubbles, followed by their expansion and eventual collision. 
We must therefore restrict attention to the parameter region in which a localized CDL bounce is the physically relevant tunneling saddle. 
A useful criterion is provided by the curvature of the potential at the top of the barrier,
\begin{equation}
H_{\rm cr}\equiv \frac{\sqrt{-\,U''(\varphi_{\mathrm{top}})}}{2}.
\end{equation}
For $H<H_{\rm cr}$, the CDL bounce typically provides the dominant tunneling channel, whereas for $H>H_{\rm cr}$ the Hawking--Moss solution becomes the relevant saddle~\cite{Sasaki:1994yj,Joti:2017fwe}. 

Since the Hawking--Moss instanton does not describe the nucleation of a bubble, but rather a homogeneous horizon-sized excursion to the top of the barrier, it is not suitable for the bubble-induced DM production mechanism considered here. 
For this reason, in the present work we impose $H<H_{\rm cr}$ 
as a basic consistency condition ensuring that the relevant tunneling channel lies in the CDL regime rather than in the Hawking--Moss regime.

\subsection{Existence and disappearance of the CDL solution}

Balek and Demetrian showed that the existence of a CDL instanton in de~Sitter space is constrained by a simple curvature criterion \cite{Balek:2003uu}. 
Let $\varphi_{\rm top}$ denote the top of the potential barrier, and let $H$ be the de~Sitter Hubble scale. 
A sufficient condition for the existence of a CDL bounce is
\begin{equation}
  4 H^2 < - U''(\varphi_{\rm top}),
\end{equation}
namely, the magnitude of the curvature at the barrier top must exceed the de~Sitter scale. 
Balek and Demetrian further showed that a necessary condition is that the inequality~\cite{Balek:2003uu}
\begin{equation}
  4 H^2 < - U''(\varphi)
\end{equation}
be satisfied at least somewhere in the barrier region. 
In general, these two conditions leave an intermediate ``grey zone'' in parameter space. 
For the quartic potential studied in Ref.~\cite{Balek:2003uu}, however, this grey zone appears to be absent, so that the top-of-the-barrier criterion effectively also becomes necessary. 
This is precisely consistent with the condition discussed above that the Hawking--Moss instanton not be the relevant decay channel.

\subsection{Nucleation-rate requirements at different inflationary epochs}

Following Turner, Weinberg, and Widrow~\cite{Turner:1992tz}, we characterize the progress of the transition by the dimensionless nucleation rate
\begin{equation}
  \gamma(N)\equiv \frac{\Gamma}{H^{4}},
\end{equation}
where $\Gamma$ is the bubble nucleation rate per unit physical four-volume, and $N$ denotes the number of $e$-folds between bubble nucleation and the end of inflation. 
It is also convenient to define
\begin{equation}
  N_{14}\equiv \ln\!\left(\frac{M}{10^{14}\,\mathrm{GeV}}\right),
  \qquad
  M\simeq \rho_{\rm inf}^{1/4}.
\end{equation}

The absence of observationally unacceptable ``big bubbles'' requires $\gamma(N)$ to remain very small during most of inflation. 
Summarizing the bounds shown in Fig.~3 of Ref.~\cite{Turner:1992tz}, one finds the approximate requirements
\begin{equation}
  \gamma(N)\lesssim
  \begin{cases}
    10^{-3}, & N\simeq 39+N_{14}, \qquad \text{(primordial nucleosynthesis)},\\[4pt]
    10^{-3}, & N\simeq 47+N_{14}, \qquad \text{($\mu$-distortion of the CMB spectrum)},\\[4pt]
    3\times 10^{-4}, & N\simeq 53+N_{14}, \qquad \text{(CMB temperature distortions)},\\[4pt]
    10^{-4}, & 50+N_{14}\lesssim N\lesssim 54+N_{14}, \qquad \text{(CMB isotropy)}.
  \end{cases}
\end{equation}

A successful completion of the phase transition, on the other hand, requires the physical false vacuum volume to begin decreasing. 
Since the false vacuum fraction is $f_{\rm fv}(t)=e^{-I(t)}$, where $I(t)$ denotes the expected volume of true vacuum bubbles per unit physical volume at time $t$, the relevant quantity is
\begin{equation}
  V_{\rm phys}(t)\propto R^{3}(t)\,f_{\rm fv}(t).
\end{equation}
The onset of percolation is then defined by
\begin{equation}
  \left.\frac{d}{dt}\!\left[R^{3}(t)f_{\rm fv}(t)\right]\right|_{t=t_e}=0,
\end{equation}
or, equivalently, \(I'(t_e)=3H\).
Turner \emph{et al.}~\cite{Turner:1992tz} showed that this implies the lower bound
\begin{equation}
  \gamma_\ast \ge \frac{9}{4\pi},
\end{equation}
which provides the percolation condition necessary for a graceful exit from first-order inflation.

In our setup, a successful FOPT during inflation must satisfy three basic requirements. 
First, bubble nucleation has to be sufficiently suppressed during the early stage of inflation in order to avoid the production of astrophysically dangerous ``big bubbles.'' 
Following Ref.~\cite{Turner:1992tz}, this requirement is expressed in terms of the dimensionless nucleation rate $\gamma(N)$, which must remain well below unity, and in practice satisfy the big bubble bounds quoted above, for bubbles nucleated long before the end of inflation. 
Of course, if the spectator field remains in the stable vacuum during this epoch, then $\Gamma= 0$, and the early-time constraint is trivially satisfied.

Second, near the end of inflation the nucleation rate must rise to an order-unity value so that the phase transition can actually complete. 
More precisely, the physical false vacuum volume must begin to decrease, which implies the percolation condition $\gamma_\ast \ge 9/(4\pi)$. 
In our numerical analysis, we implement this requirement conservatively as
\begin{equation}
  \gamma_\ast \ge 1.
\end{equation}

Third, in addition to percolation, the phase transition must be sufficiently rapid for bubbles to overlap and collide efficiently before inflationary expansion separates them. 
To ensure this, we impose the conservative requirement
\begin{equation}
  \frac{\beta}{H} \gtrsim 10.
\end{equation}
Here $\beta$ characterizes the inverse time scale of the phase transition and is defined by
\begin{equation}
  \beta
  \equiv
  \left.\frac{1}{\Gamma}\frac{d\Gamma}{dt}\right|_{t=t_\star}
  \simeq
  -\left.\frac{dS_E}{dt}\right|_{t=t_\star},
\end{equation}
where $\Gamma$ is the bubble nucleation rate per unit volume, $S_E$ is the Euclidean tunneling action, and $t_\star$ denotes the characteristic time of the transition. 
In this work, we choose $t_\star$ to be the moment at which $\gamma(N)=1$. 
This choice is stronger than the minimal requirement obtained from demanding that bubble collisions occur within one $e$-fold, and its order of magnitude is consistent with the estimate adopted in Ref.~\cite{An:2020fff}. 
Indeed, following Ref.~\cite{Zou:2026wzi}, the collision condition can be written as
\begin{equation}
  d_{\rm sep}(t_i)\le \frac{1-e^{N-N_i}}{H},
\end{equation}
where $d_{\rm sep}(t_i)$ denotes the average initial physical separation between bubbles at the nucleation time $t_i$, while $N_i$ and $N$ are the corresponding e-folding numbers at $t_i$ and at the collision time $t$, respectively. 
For a collision to occur within one e-fold after nucleation, namely $N_i-N=1$, this condition becomes
\begin{equation}
  d_{\rm sep}(t_i)\le \frac{1-e^{-1}}{H}.
\end{equation}
Estimating the typical bubble separation by the characteristic collision radius, \(d_{\rm sep}(t_i)\sim R_{\rm col}\sim v_w/\beta\), one finds
\begin{equation}
  \frac{\beta}{H}\gtrsim \frac{v_w}{1-e^{-1}}
  \simeq 1.58\,v_w.
\end{equation}
Therefore, for relativistic bubble walls with \(v_w \simeq 1\), the condition \(\beta/H \gtrsim \mathcal{O}(1)\) is already sufficient to ensure bubble overlap within one \(e\)-fold. 
Our benchmark choice $\beta/H \gtrsim 10$ therefore ensures that this condition is well satisfied and that the phase transition is not only able to complete, but also sufficiently short-lived to make bubble collisions efficient and to justify treating the transition as a relatively sharp event near the end of inflation.

\begin{figure}[h]
	\centering
	\includegraphics[width=0.8\textwidth]{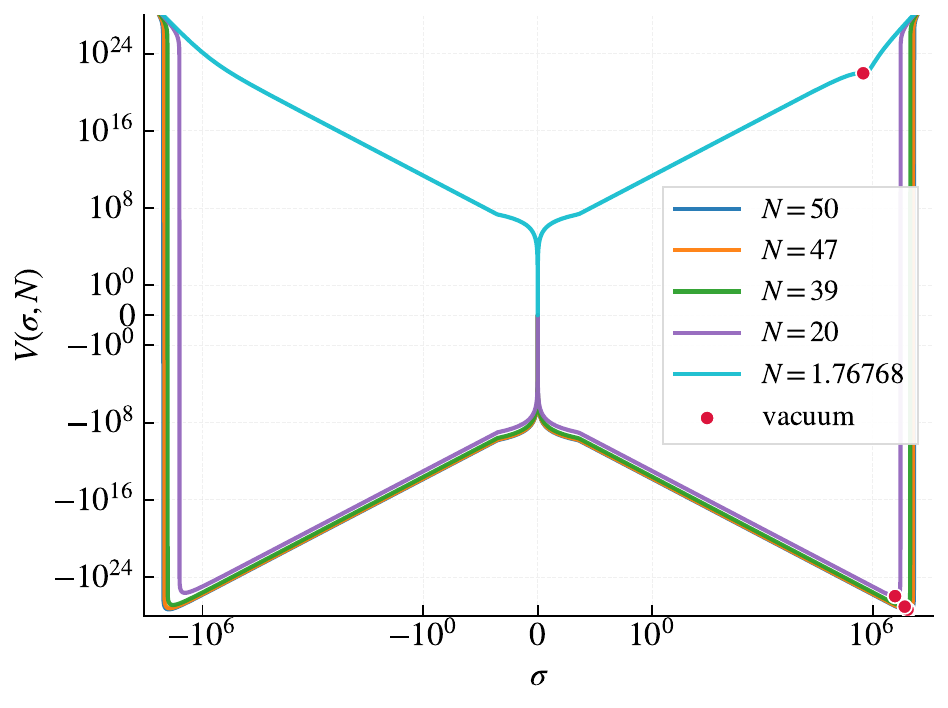}
    \caption{Evolution of the spectator-field potential with the inflationary $e$-fold number $N$. Both axes are shown on a symmetric logarithmic scale. The red markers indicate the vacuum continuously occupied by the spectator field: it corresponds to the true vacuum at early times, but is lifted to a false vacuum near the end of inflation, thereby triggering the onset of the phase transition.}
	\label{figvaccum}
\end{figure}

In addition, to ensure that the phase transition sector does not significantly disturb the inflationary background, we require the change in vacuum energy along the branch continuously occupied by the spectator field---namely, the red points in Fig.~\ref{figvaccum}---to remain subdominant compared with the inflationary energy density. 
More explicitly, if $\sigma_{\rm occ}(N)$ denotes the vacuum continuously followed by the spectator field, we impose
\begin{equation}
  \left|\,V\bigl(\sigma_{\rm occ}(N_*),N_*\bigr)-V\bigl(\sigma_{\rm occ}(60),60\bigr)\right|
  \;<\; 0.1\,\rho_{\rm inf}.
\label{occupequ}
\end{equation}
Here $N=60$ is taken as a benchmark reference point for the beginning of the inflationary stage relevant to our analysis, while $N=N_*$ denotes the time of the phase transition. 
The numerical factor $0.1$ is not intended as a sharp theoretical boundary, but rather as a simple benchmark to ensure that the spectator sector remains subdominant and that the phase transition does not induce significant backreaction on the inflationary evolution~\cite{Bao:2026bgu,An:2020fff}.
A more refined treatment of the backreaction is left for future work.

As illustrated in Fig.~\ref{figvaccum}, for the parameter points that survive our final selection, the spectator field initially sits stably in the vacuum that is the true minimum of the potential during the early stage of inflation. 
In this regime, no phase transition takes place and the nucleation rate is effectively absent. 
As inflation proceeds toward its end, the potential is gradually lifted and the vacuum continuously occupied by the spectator field becomes metastable, namely, it evolves from the true vacuum into a false vacuum. 
Only then does vacuum decay become possible and the phase transition begin. 
This late-time onset of metastability naturally prevents the formation of excessively large bubbles during the earlier stage of inflation.

\subsection{Computational methodology for the nucleation rate}

Following Sasaki, Stewart, and Tanaka (SST)~\cite{Sasaki:1994yj}, we rewrite the spectator potential at fixed inflaton background in the quartic form used in their analysis of vacuum decay in de~Sitter space. 
Since throughout this work we fix $\lambda=-1$, the spectator potential is
\begin{equation}
V(\sigma;\phi)
=
\frac12\,m^2(\phi)\,\sigma^2
-\frac{\epsilon}{3}\,\sigma^3
+\frac{\kappa}{4}\,\sigma^4,
\qquad
m^2(\phi)=\mu^2-c_{\rm inf}^2\phi^2,
\label{eq:Vsigma_fixedphi}
\end{equation}
with $\kappa>0$. 
Therefore, besides $\sigma=0$, the nontrivial extrema are
\begin{equation}
\sigma_\pm(\phi)
=
\frac{\epsilon\pm\sqrt{\Delta(\phi)}}{2\kappa},
\qquad
\Delta(\phi)\equiv \epsilon^2-4\kappa\,m^2(\phi).
\end{equation}
In the metastable regime relevant for our analysis, we identify
\begin{equation}
\sigma_{\rm tv}=0,
\qquad
\sigma_{\rm fv}=\sigma_+(\phi),
\qquad
V(\sigma_{\rm fv};\phi)>V(\sigma_{\rm tv};\phi)=0.
\end{equation}

To match the SST convention, in which the false vacuum is located at the origin, we introduce the shifted spectator field
\begin{equation}
\varphi_{\mathrm{PT}} \equiv \sigma_{\rm fv}(\phi)-\sigma .
\end{equation}
Then $\varphi_{\mathrm{PT}}=0$ corresponds to the false vacuum, while $\varphi_{\mathrm{PT}}=\sigma_{\rm fv}(\phi)$ corresponds to the true vacuum. 
Expanding the potential around $\sigma=\sigma_{\rm fv}(\phi)$ and using the stationarity condition
\begin{equation}
m^2(\phi)\,\sigma_{\rm fv}
-\epsilon\,\sigma_{\rm fv}^{\,2}
+\kappa\,\sigma_{\rm fv}^{\,3}=0,
\end{equation}
the linear term vanishes identically, and the potential can be written exactly as
\begin{equation}
V(\varphi;\phi)
=
V_0(\phi)
+\frac12\,m_{\rm SST}^2(\phi)\,\varphi_{\mathrm{PT}}^2
-\mu_3(\phi)\,\varphi_{\mathrm{PT}}^3
+\lambda_4\,\varphi_{\mathrm{PT}}^4,
\label{eq:V_SST_mapping}
\end{equation}
where
\begin{align}
\mu_3(\phi)
&=
\frac{\epsilon+3\sqrt{\Delta(\phi)}}{6},\\
m_{\rm SST}^2(\phi)
&=
\frac{\Delta(\phi)+\epsilon\sqrt{\Delta(\phi)}}{2\kappa}
=
\frac{\sqrt{\Delta(\phi)}\bigl(\sqrt{\Delta(\phi)}+\epsilon\bigr)}{2\kappa},\\
V_0(\phi)
&=
V\!\left(\sigma_{\rm fv}(\phi);\phi\right)
=
\frac{\bigl(\epsilon+\sqrt{\Delta(\phi)}\bigr)^3\bigl(\epsilon-3\sqrt{\Delta(\phi)}\bigr)}
{192\,\kappa^3},\\
\lambda_4&=\frac{\kappa}{4}.
\end{align}
Here \(V_0(\phi)\) represents only the spectator-sector false vacuum energy and excludes the inflaton potential \(V_{\rm inf}(\phi)\).
In the negligible-backreaction regime, the bounce action depends only on the dimensionless combinations introduced by SST,
\begin{equation}
H_0(\phi)\equiv \frac{H}{m_{\rm SST}(\phi)},
\qquad
\nu(\phi)\equiv \frac{2\lambda_4\,m_{\rm SST}^2(\phi)}{\mu_3^2(\phi)}
=
\frac{\kappa\,m_{\rm SST}^2(\phi)}{2\,\mu_3^2(\phi)},
\end{equation}
together with
\begin{equation}
\epsilon_{\rm SST}(\phi)\equiv 1-\nu(\phi),
\qquad
\delta(\phi)\equiv \sqrt{\epsilon_{\rm SST}^2(\phi)+H_0^2(\phi)}.
\end{equation}
The SST fitting formula for the bounce exponent is then
\begin{equation}
B(\phi;H)
=
\frac{m_{\rm SST}^2(\phi)\,\pi^2}{12\,\mu_3^2(\phi)}\,
\widetilde B\!\bigl(\nu(\phi),H_0(\phi)\bigr),
\label{eq:Bphi_final}
\end{equation}
where, for $0\le \nu\le 1$,
\begin{align}
\widetilde B
&=
\frac{4\bigl(1+10.07\,\epsilon_{\rm SST}+16.55\,\epsilon_{\rm SST}^2\bigr)}
{\delta\bigl(\epsilon_{\rm SST}+\delta\bigr)^2(1+10\,\epsilon_{\rm SST})}
\left[
1+\frac{11}{2}\epsilon_{\rm SST}
+3\frac{\epsilon_{\rm SST}^2}{\delta}
+\frac{3}{2}\frac{\epsilon_{\rm SST}^3}{\delta^2}
\right]\widetilde B_1,
\\
\widetilde B_1
&=
1+\sum_{i,j} k_{ij}\,\epsilon_{\rm SST}^{\,i}\,\delta^{\,j},
\end{align}
with nonvanishing coefficients
\begin{align}
&k_{01}=-0.1617,\qquad k_{03}=-5.507,\nonumber\\
&k_{11}=-11.34,\qquad k_{12}=30.17,\qquad k_{14}=-21.69,\nonumber\\
&k_{20}=12.09,\qquad k_{21}=-33.24,\qquad k_{23}=41.29,\nonumber\\
&k_{30}=7.728,\qquad k_{32}=-19.34.
\end{align}

The SST fitting formula is applicable only when gravitational backreaction is negligible. 
In our notation, this requires
\begin{equation}
\kappa_{\rm SST}(\phi)
\equiv
\frac{8\pi G\,m_{\rm SST}^4(\phi)}{\mu_3^2(\phi)}
\ll
\min\!\Bigl\{1,\max\bigl(H_0(\phi),\epsilon_{\rm SST}(\phi)\bigr)\Bigr\}.
\end{equation}
Outside this regime, the SST fit is unreliable, and the gravitational backreaction should be treated explicitly.

Finally, we estimate the nucleation rate as
\begin{equation}
\Gamma \simeq m_{\rm char}^4\,e^{-B},
\end{equation}
where \(m_{\rm char}\) denotes the characteristic mass scale of the spectator sector. 
In practice, we take the prefactor to be of order \(m^4\). 
It then follows that
\begin{equation}
\frac{\beta}{H}
=
\left.\frac{1}{\Gamma}\frac{d\Gamma}{dt}\right/ H
=
-\frac{d\ln \Gamma}{dN}
\approx
\frac{dB}{dN}.
\label{eq:beta_over_H_final}
\end{equation}
Therefore, once \(B(\phi;H)\) is determined from Eq.~\eqref{eq:Bphi_final}, the quantity \(\beta/H\) can be obtained directly from the variation of the bounce action with respect to the number of remaining \(e\)-folds.

\subsection{Summary of conditions used in the analysis}

For convenience, we summarize in Table~\ref{tab:vacuum-decay-conditions} the main criteria adopted in our analysis of vacuum decay and bubble nucleation during inflation. 
Together, these criteria define the parameter region in which the phase transition admits a physically meaningful localized bubble interpretation, remains compatible with the inflationary background, and completes efficiently near the end of inflation.

\begin{table}[t]
\centering
\small
\setlength{\tabcolsep}{5pt}
\renewcommand{\arraystretch}{1.15}
\caption{Summary of the main criteria adopted in the analysis.}
\label{tab:vacuum-decay-conditions}
\begin{tabularx}{\textwidth}{@{}p{3.1cm} p{4.8cm} >{\raggedright\arraybackslash}X@{}}
\hline
Criterion & Expression used & Purpose \\
\hline

\multicolumn{3}{@{}l}{\textit{Bounce interpretation and tunneling channel}} \\

Small type-A regime 
& \(\eta \ll M_{\rm Pl}\), with \(\Delta U\) not parametrically small 
& Restricts the analysis to the small type-A regime, so that the CDL bounce retains the standard semiclassical interpretation and the additional negative modes are avoided. \\
\hline
CDL regime 
& \(H < \sqrt{-U''(\phi_{\rm top})}/2\)
& Ensures that a localized CDL bounce exists and remains the physically relevant tunneling channel, rather than the Hawking--Moss transition. \\
\hline
\multicolumn{3}{@{}l}{\textit{Cosmological viability of the transition}} \\

Big-bubble bound (BBN) 
& \(\gamma(N) \lesssim 10^{-3}\) at \(N \simeq 39 + N_{14}\)
& Prevents the production of excessively large bubbles that would conflict with primordial-nucleosynthesis constraints. \\
\hline
Big-bubble bound (\(\mu\)-distortion) 
& \(\gamma(N) \lesssim 10^{-3}\) at \(N \simeq 47 + N_{14}\)
& Prevents early bubble production from generating an unacceptable \(\mu\)-distortion of the CMB spectrum. \\
\hline
Big-bubble bound (CMB temperature distortions) 
& \(\gamma(N) \lesssim 3 \times 10^{-4}\) at \(N \simeq 53 + N_{14}\)
& Prevents early bubble production from generating excessive CMB temperature distortions. \\
\hline
Big-bubble bound (CMB isotropy) 
& \(\gamma(N) \lesssim 10^{-4}\) for \(50 + N_{14} \lesssim N \lesssim 54 + N_{14}\)
& Ensures consistency with the observed large-scale isotropy of the CMB. \\
\hline
Percolation requirement 
& \(\gamma_* \ge 1\) 
& Requires the nucleation rate to become large enough near the end of inflation for the phase transition to complete. \\
\hline
Rapid-transition requirement 
& \(\beta/H \gtrsim 10\) 
& Ensures that bubbles overlap and collide efficiently before inflationary expansion separates them. \\
\hline
Subdominant backreaction 
& \begin{tabular}[t]{@{}l@{}}
\(\bigl|V(\sigma_{\rm occ}(N_*),N_*)\) \\
\(\qquad -\,V(\sigma_{\rm occ}(60),60)\bigr| < 0.1\,\rho_{\rm inf}\)
\end{tabular}
& Requires the vacuum-energy variation along the continuously occupied branch of the spectator field to remain subdominant relative to the inflationary background. \\

\hline
\end{tabularx}

\end{table}

These criteria will be imposed jointly in the numerical analysis below to identify the viable parameter region relevant for DM production from bubble collisions.

\section{DM production from bubble collisions}
\label{sec:DMproduction}
Having established the conditions under which the phase transition can proceed through physically meaningful bubble nucleation during inflation, we now turn to the resulting particle production from bubble collisions and its impact on the present-day DM abundance. 
Our analysis is not restricted to direct DM production alone, but instead incorporates the full set of relevant production channels arising from bubble collisions, which can in general excite all fields coupled to the spectator field, and follows the subsequent post-collision evolution of the produced populations.
In particular, besides the direct production of dark fermions, we also take into account the production of spectator field and inflaton excitations, together with their later decays, scatterings, and cosmological dilution. 
This allows us to determine the final DM abundance in a more complete way than an analysis based only on the direct collision-induced DM yield.

\subsection{Particle production from bubble collisions}

Bubble collisions release part of the energy stored in the bubble walls and can thereby excite particles. 
For the quartic potential considered here, we denote the false vacuum energy density, true vacuum energy density, and barrier top potential value by \(V_{\rm fv}\), \(V_{\rm tv}\), and \(V_{\rm top}\), respectively. 
A useful quantity for describing the post-collision evolution is the collision degeneracy parameter~\cite{Mansour:2023fwj,Jinno:2019bxw},
\begin{equation}
\varepsilon_{\rm coll}
\equiv
\frac{V_{\rm top}-V_{\rm fv}}{V_{\rm top}-V_{\rm tv}} .
\end{equation}
For this class of potentials, \(\varepsilon_{\rm coll}\) serves as an effective indicator of the collision outcome. 
In particular, the regime
\begin{equation}
\varepsilon_{\rm coll}>0.22
\end{equation}
corresponds to elastic collisions, whereas
\begin{equation}
\varepsilon_{\rm coll}<0.22
\end{equation}
corresponds to inelastic ones. 
In the elastic regime, the field excursion after collision is sufficient to drive the configuration back across the barrier, so that the region between the outgoing walls evolves again toward the false vacuum side. 
In the inelastic regime, this restoration does not occur; instead, the released energy is deposited into localized oscillations of the scalar field on the true vacuum side. 
In either case, the subsequent dynamics involve oscillatory motion of the scalar field around the vacuum selected by the post-collision configuration.

To compute particle production, one decomposes the time-dependent background field configuration into Fourier modes labeled by the four-momentum
\(
p^\mu = (\omega,\mathbf{k})
\),
with \(\omega\) the frequency and \(\mathbf{k}\) the spatial momentum. Its Lorentz-invariant square is
\(
p^2 = \omega^2 - k^2
\),
where \(k \equiv |\mathbf{k}|\). The particle yield is then related to the imaginary part of the two-point 1PI function.
The number $N$ and energy $E$ of produced particles per unit bubble wall area $A$ are then~\cite{Mansour:2023fwj}
\begin{align}
\frac{N}{A}
&=
\frac{1}{4\pi^2}
\int_{p_{\min}^2}^{p_{\max}^2}
dp^2\,
f(p^2)\,
\operatorname{Im}\Gamma^{(2)}(p^2),
\\
\frac{E}{A}
&=
\frac{1}{8\pi^2}
\int_{p_{\min}^2}^{p_{\max}^2}
dp^2\,
f(p^2)\,
\sqrt{p^2}\,
\operatorname{Im}\Gamma^{(2)}(p^2).
\end{align}
Here $f(p^2)$ is the efficiency factor that encodes the Fourier-space structure of the bubble collision, while $\operatorname{Im}\Gamma^{(2)}(p^2)$ contains the model-dependent probability for the off-shell scalar excitation to decay into the final states of interest.  
The number density and energy density are then estimated as
\begin{equation}
n \sim \frac{1}{R_*}\,\frac{N}{A},
\qquad
\rho \sim \frac{1}{R_*}\,\frac{E}{A}.
\end{equation}
This provides an order-of-magnitude prescription for relating the particle production per unit wall area to the initial number and energy densities after bubble collisions, where $R_*$ denotes the typical bubble radius at the time of collision. 
In our estimates, we take
\begin{equation}
R_*\sim \frac{v_w}{\beta} .
\end{equation}
The infrared cutoff is set by the larger of the physical production threshold and the inverse bubble size,
\begin{equation}
p_{\min}\sim \max\!\left(m_{\rm thr},R_*^{-1}\right).
\end{equation}
Here \(m_{\rm thr}\) denotes the kinematic mass threshold associated with the produced final state particles. The naive ultraviolet cutoff is determined by the boosted wall thickness,
\begin{equation}
p_{\max}^2=\left(\frac{\gamma_w}{l_w}\right)^2.
\end{equation}
The cutoffs \(p_{\min}\) and \(p_{\max}\) specify the finite momentum range set by the physical scales of the bubble configuration. Modes with momenta below \(p_{\min}\) are either kinematically inaccessible or sensitive to multi-bubble effects on scales larger than the characteristic bubble size \(R_*\). In the ultraviolet region, they are not applicable at distances shorter than the boosted wall thickness \(l_w/\gamma_w\), since finite-width effects of the bubble wall, which are not captured here, become important~\cite{Shakya:2023kjf}. In this work, following the simplified treatment used in our numerical estimates, we fix the wall Lorentz factor to the benchmark value
\begin{equation}
\gamma_w = 200,
\end{equation}
and estimate the wall thickness as
\begin{equation}
l_w \sim v_\sigma^{-1},
\end{equation}
where $v_\sigma$ denotes the field value at the false vacuum.

For a perfectly elastic collision, the efficiency factor is known analytically~\cite{Mansour:2023fwj}:
\begin{equation}
f_{\rm PE}(p^2)
=
\frac{16v_\sigma^2}{p^4}
\log\!\left[
\frac{
2(\gamma_w/l_w)^2-p^2+2(\gamma_w/l_w)\sqrt{(\gamma_w/l_w)^2-p^2}
}{
p^2
}
\right]
\Theta\!\left[(\gamma_w/l_w)^2-p^2\right].
\end{equation}
A key result of Ref.~\cite{Mansour:2023fwj} is that both elastic and inelastic collisions contain the same universal ultraviolet contribution, encoded by the perfectly elastic efficiency factor \(f_{\rm PE}(p^2)\), which arises from the short distance collision of the bubble walls. The difference between the two cases is mainly an additional enhancement induced by the post-collision scalar oscillations.

A convenient fit to the numerical results is~\cite{Mansour:2023fwj}
\begin{align}
f_{\rm elastic}(p^2)
&=
f_{\rm PE}(p^2)
+
\frac{v_\sigma^2L_p^2}{15m_t^2}
\exp\!\left[
-\frac{\bigl(p^2-m_t^2+12m_t/L_p\bigr)^2}{440\,m_t^2/L_p^2}
\right],
\\
f_{\rm inelastic}(p^2)
&=
f_{\rm PE}(p^2)
+
\frac{v_\sigma^2L_p^2}{4m_f^2}
\exp\!\left[
-\frac{\bigl(p^2-m_f^2+31m_f/L_p\bigr)^2}{650\,m_f^2/L_p^2}
\right].
\end{align}
Here $f_{\rm PE}(p^2)$ captures the universal power-law contribution associated with the bubble collision itself, while the Gaussian term parameterizes the peak generated by the post-collision oscillations of the scalar field.
The characteristic scale entering the oscillatory contribution is
\begin{equation}
L_p \equiv \min\!\left(R_*,\Gamma_\sigma^{-1}\right),
\end{equation}
where $\Gamma_\sigma$ denotes the decay width of the spectator field excitation. 
The effective scalar masses around the false and true vacua are
\begin{equation}
m_f^2=
\left.\frac{d^2V(\sigma)}{d\sigma^2}\right|_{\sigma_{\rm fv}},
\qquad
m_t^2=
\left.\frac{d^2V(\sigma)}{d\sigma^2}\right|_{\sigma_{\rm tv}}.
\end{equation}
These fit functions therefore provide a compact parameterization of both the universal ultraviolet contribution from the bubble wall collision and the collision-dependent enhancement associated with the subsequent scalar oscillations~\cite{Mansour:2023fwj,Cataldi:2025nac}.

In our numerical parameter space, we find that the bubble collisions are always inelastic. 
In this case, the post-collision field configuration evolves on the true vacuum side and subsequently oscillates around the true vacuum minimum.
As discussed in Ref.~\cite{Shakya:2023kjf}, the oscillation-induced contribution to particle production should be evaluated using the masses and interactions associated with the vacuum around which the field oscillates. 
Although the description of the collision stage can in general be subtle when the field explores configurations interpolating between the two vacua during the impact, the relevant late-time background in our case is unambiguously the true vacuum configuration. 
We therefore compute the particle production using the masses and interactions defined in the true vacuum.

For notational clarity, we denote by \(h\) the fluctuation of the spectator field around the post-transition true vacuum, chosen to be at \(\sigma=0\). 
Writing the inflaton as
\begin{equation}
\phi=\bar{\phi}+\varphi_{\rm{inf}},
\end{equation}
the relevant part of the Lagrangian can be written as
\begin{equation}
\begin{aligned}
\mathcal{L}\supset\;&
\frac{1}{2}\partial_\mu h\,\partial^\mu h
+\frac{1}{2}\partial_\mu \varphi_{\rm inf}\,\partial^\mu \varphi_{\rm inf}
+\bar{\psi}\,i\gamma^\mu\partial_\mu\psi  \\
&-\frac{1}{2}m_h^2 h^2
-\frac{\lambda\epsilon}{3}h^3
-\frac{\kappa}{4}h^4
-y\,h\,\bar{\psi}\psi  \\
&+c_{\rm inf}^2\bar{\phi}\,\varphi_{\rm inf}\,h^2
+\frac{c_{\rm inf}^2}{2}\varphi_{\rm inf}^2 h^2 .
\end{aligned}
\label{interaction}
\end{equation}
where
\begin{equation}
m_h^2 \equiv \mu^2-c_{\rm inf}^2\bar{\phi}^{\,2}
\end{equation}
is the effective mass squared of the spectator field in the true vacuum background. 
Therefore, the total contribution entering the production formula is obtained by summing over all open channels,
\begin{equation}
\operatorname{Im}\Gamma^{(2)}(p^2)
=
\sum_X \operatorname{Im}\Gamma^{(2)}_{h^\ast\to X}(p^2),
\qquad
X\in\{hh,\;3h,\;h\varphi_{\rm{inf}},\;h\varphi_{\rm{inf}}\varphi_{\rm{inf}},\;\bar\psi\psi\}.
\end{equation}
For the phase-transition scalar final states, one has
\begin{align}
\operatorname{Im}\Gamma^{(2)}_{h^\ast\to hh}(p^2)
&=
\frac{\lambda^2\epsilon^2}{8\pi}
\sqrt{1-\frac{4m_h^2}{p^2}}\,
\Theta(p^2-4m_h^2),
\\
\operatorname{Im}\Gamma^{(2)}_{h^\ast\to 3h}(p^2)
&=
\frac{3\kappa^2 p^2}{256\pi^3}
\sqrt{1-\frac{9m_h^2}{p^2}}\,
\Theta(p^2-9m_h^2).
\end{align}
For the inflaton-related channels, we take \(m_{\varphi_{\rm{inf}}} \sim m_h \sim m\) as a representative simplifying choice. This approximation is adequate for estimating the corresponding yields at the order-of-magnitude level. If the inflaton is heavier than the phase-transition field, \(m_{\varphi_{\rm{inf}}} > m_h\), the corresponding yield is further suppressed, because the lower bounds of the \(N/A\) and \(E/A\) integrals are shifted to larger values, while \(f_{\rm PE}\) is reduced at the same time. This makes the inflaton-related contribution even less important and therefore further justifies the later simplification of the Boltzmann system. If instead the inflaton is lighter, \(m_{\varphi_{\rm{inf}}} < m_h\), the presence of the final state \(h\) particle ensures that the lower bounds of the \(N/A\) and \(E/A\) integral remain at the same order, so the resulting yield is not expected to differ significantly from the estimate obtained in the equal-mass case. We therefore conclude that the choice \(m_{\varphi_{\rm{inf}}} \sim m_h \sim m\) provides a reasonable and robust approximation for our purposes. In addition, in our simplified treatment the inflaton--spectator coupling \(c_{\rm inf}\) is assumed to be extremely small. As a result, the precise choice of \(m_{\varphi_{\rm{inf}}}\) in this range has only a negligible effect on the final result. More explicitly, the corresponding imaginary parts are given by
\begin{align}
\operatorname{Im}\Gamma^{(2)}_{h^\ast\to h\varphi_{\rm{inf}}}(p^2)
&=
\frac{c_{\rm inf}^4\bar\phi^2}{4\pi}
\sqrt{1-\frac{4m^2}{p^2}}\,
\Theta(p^2-4m^2),
\\
\operatorname{Im}\Gamma^{(2)}_{h^\ast\to h\varphi_{\rm{inf}}\varphi_{\rm{inf}}}(p^2)
&=
\frac{c_{\rm inf}^4 p^2}{256\pi^3}
\sqrt{1-\frac{9m^2}{p^2}}\,
\Theta(p^2-9m^2).
\end{align}
Finally, for fermionic DM,
\begin{equation}
\operatorname{Im}\Gamma^{(2)}_{h^\ast\to\bar\psi\psi}(p^2)
=
\frac{y^2p^2}{8\pi}
\left(1-\frac{4m_\psi^2}{p^2}\right)^{3/2}
\Theta(p^2-4m_\psi^2).
\end{equation}
These expressions make it manifest that bubble collisions provide a unified non-thermal source for the production of dark fermions \(\psi\), spectator particles \(h\), and inflaton excitations \(\varphi_{\rm{inf}}\).

A final issue is the validity of the ultraviolet cutoff. If the energy injected into produced particles becomes comparable to the vacuum energy released by the phase transition, particle production backreacts on the wall profile and the naive estimate \(p_{\max}=\gamma_w/l_w\) may overestimate the yield. Following Ref.~\cite{Cataldi:2025nac}, we account for this effect by introducing a self-consistent cutoff \(\mu_M\), defined schematically by
\begin{equation}
\sum_i \rho_i(\mu_M)\sim \Delta V,
\end{equation}
and replacing the upper integration bound by
\begin{equation}
p_{\max}^2
=
\min\!\left[
\left(\frac{\gamma_w}{l_w}\right)^2,\,
\mu_M^2
\right].
\end{equation}
The physical interpretation is that once particle emission becomes efficient enough to significantly drain energy from the wall, the collision backreacts on the wall structure and effectively increases its thickness, thereby lowering the maximum momentum available in the Fourier decomposition. In practice, this yields a more conservative estimate of both \(N/A\) and \(E/A\), while leaving the formalism otherwise unchanged.

For the phenomenological analysis below, we therefore adopt the prescription
\begin{equation}
f(p^2)=f_{\rm inelastic}(p^2),
\qquad
p_{\max}^2=
\min\!\left[
\left(\frac{\gamma_w}{l_w}\right)^2,\,
\mu_M^2
\right],
\qquad
p_{\min}^2=
\max\!\left(
m_{\rm thr}^2,\,
R_*^{-2}
\right).
\end{equation}
together with the full model-dependent ${\rm Im}\,\Gamma^{(2)}(p^2)$ for all kinematically open channels.

\subsection{Subsequent evolution of particles produced in bubble collisions}

As discussed in the previous subsection, bubble collisions can produce three species in our setup: the spectator field \(h\), the inflaton fluctuations \(\varphi_{\rm{inf}}\), and the DM fermions \(\psi\). After their initial production, these species continue to evolve through decays or scatterings induced by the interactions in Eq.~\eqref{interaction}. In  general, their number densities satisfy a coupled system of Boltzmann equations,
\begin{align}
a^{-3}\frac{d\!\left(n_h a^3\right)}{dt}
&=
S_h^{\rm coll}
+
\mathcal{C}_h^{\rm tot},
\\
a^{-3}\frac{d\!\left(n_\psi a^3\right)}{dt}
&=
S_\psi^{\rm coll}
+
\mathcal{C}_\psi^{\rm tot},
\\
a^{-3}\frac{d\!\left(n_{\varphi_{\rm{inf}}} a^3\right)}{dt}
&=
S_{\varphi_{\rm{inf}}}^{\rm coll}
+
\mathcal{C}_{\varphi_{\rm{inf}}}^{\rm tot},
\end{align}
where \(S_i^{\rm coll}\) denote the source terms associated with particle production from bubble collisions, while \(\mathcal{C}_i^{\rm tot}\) denotes the total collision term for species \(i\), collectively encoding the subsequent processes relevant to that species, such as scatterings and, when allowed, decays, which may differ among \(h\), \(\psi\), and \(\varphi_{\rm inf}\).

A complete solution of this coupled system is unnecessarily complicated for our present purpose. Instead, we focus on a restricted but phenomenologically relevant region of parameter space in which the late-time evolution simplifies considerably. First, we take the inflaton--spectator coupling \(c_{\rm inf}\) to be sufficiently small. In this regime, the production of inflaton particles \(\varphi_{\rm{inf}}\) from bubble collisions is strongly suppressed compared with the production of \(h\) and \(\psi\), and the subsequent processes involving \(\varphi_{\rm{inf}}\) are likewise negligible. We may therefore consistently neglect the inflaton population in the Boltzmann evolution.

Next, we further restrict the parameter space such that the number-changing scatterings
\begin{equation}
3h \to 2h,
\qquad
hh \to \bar\psi\psi,
\qquad
hh \to h\varphi_{\rm{inf}},
\qquad
hh \to \varphi_{\rm{inf}}\varphi_{\rm{inf}},
\end{equation}
remain inefficient throughout the evolution. 
More precisely, for each process \(i\), we require the corresponding interaction rate per particle to satisfy
\begin{equation}
\frac{\Gamma_i}{H}\ll 1,
\end{equation}
where \(\Gamma_i\) denotes the reaction rate associated with that process, estimated using the relevant post-collision number densities. 
Schematically, for the \(3h\to 2h\) process one has
\begin{equation}
\Gamma_{3\to2}\sim n_h^2\langle \sigma v^2\rangle,
\end{equation}
whereas for the \(2\to2\) processes one has
\begin{equation}
\Gamma_{2\to2}\sim n_h\langle \sigma v\rangle.
\end{equation}
Under this condition, all other scattering processes can be neglected. The late-time dynamics are then dominated by \(hh\to hh\), which drives kinetic equilibration within the spectator sector, together with the decay \(h\to \bar{\psi}\psi\), which transfers the spectator-sector energy into DM, since the relevant interaction rates satisfy \(\Gamma/H> 1\). The Boltzmann equations then reduce to
\begin{align}
a^{-3}\frac{d\!\left(n_h a^3\right)}{dt}
&=
S_h^{\rm coll}
-
\Gamma_{h\to\bar\psi\psi}\,n_h,
\\
a^{-3}\frac{d\!\left(n_\psi a^3\right)}{dt}
&=
S_\psi^{\rm coll}
+
2\,\Gamma_{h\to\bar\psi\psi}\,n_h,
\\
a^{-3}\frac{d\!\left(n_{\varphi_{\rm{inf}}} a^3\right)}{dt}
&\simeq
0,
\end{align}
where \(\Gamma_{h\to\bar\psi\psi}\) is the decay rate of \(h\) into a DM pair. 

Under the above assumptions, the late-time evolution becomes particularly simple. 
The only relevant process after bubble collisions is the decay \(h\to \bar\psi\psi\). 
As a result, each spectator particle \(h\) eventually produces two DM particles. 
Therefore, the final comoving DM number is given by the DM particles produced directly from bubble collisions plus those generated subsequently from the decay of \(h\).

Denoting by \(n_\psi^{\rm coll}\) and \(n_h^{\rm coll}\) the number densities of \(\psi\) and \(h\) produced at the time of bubble collisions, the corresponding DM energy density today is
\begin{equation}
\rho_{\psi,0}
=
m_\psi
\left(
n_\psi^{\rm coll}+2\,n_h^{\rm coll}
\right)
e^{-3\left(N_{\ast}+N_{\rm today}\right)},
\label{eq:rhoDMtoday_match}
\end{equation}
where the factor \(e^{-3(N_{\ast}+N_{\rm today})}\) accounts for the dilution of the number density due to the cosmic expansion from the time of bubble collisions to the present epoch. We define \(N_* \equiv \ln\!\left(a_{\rm Rh}/a_*\right)\) and \(N_{\rm today} \equiv \ln\!\left(a_0/a_{\rm Rh}\right)\), which correspond to the number of \(e\)-folds from the phase transition to the end of inflation and from the end of inflation to today, respectively.

In our analysis, we require the final DM energy density to reproduce the observed value today~\cite{Planck:2018nkj,Planck:2018lbu},
\begin{equation}
\rho_{\psi,0}
=
\rho_{\rm DM,0}
\simeq
9.6\times10^{-48}\ {\rm GeV}^4.
\label{eq:DMmatching}
\end{equation}
We then fix
\begin{equation}
y=0.1,
\end{equation}
as a perturbative benchmark large enough to make \(h\to\bar\psi\psi\) efficient in the selected parameter region, and determine the DM mass \(m_\psi\) by imposing Eq.~\eqref{eq:DMmatching}. The explicit expressions for the relevant scattering cross sections are collected in Appendix~\ref{app:cross_sections}. The numerical results obtained in this work cover only part of the parameter space under the simplifying assumptions adopted here, since a more complete exploration is currently limited by computational resources.

\begin{figure}[h]
	\centering
	\includegraphics[width=0.8\textwidth]{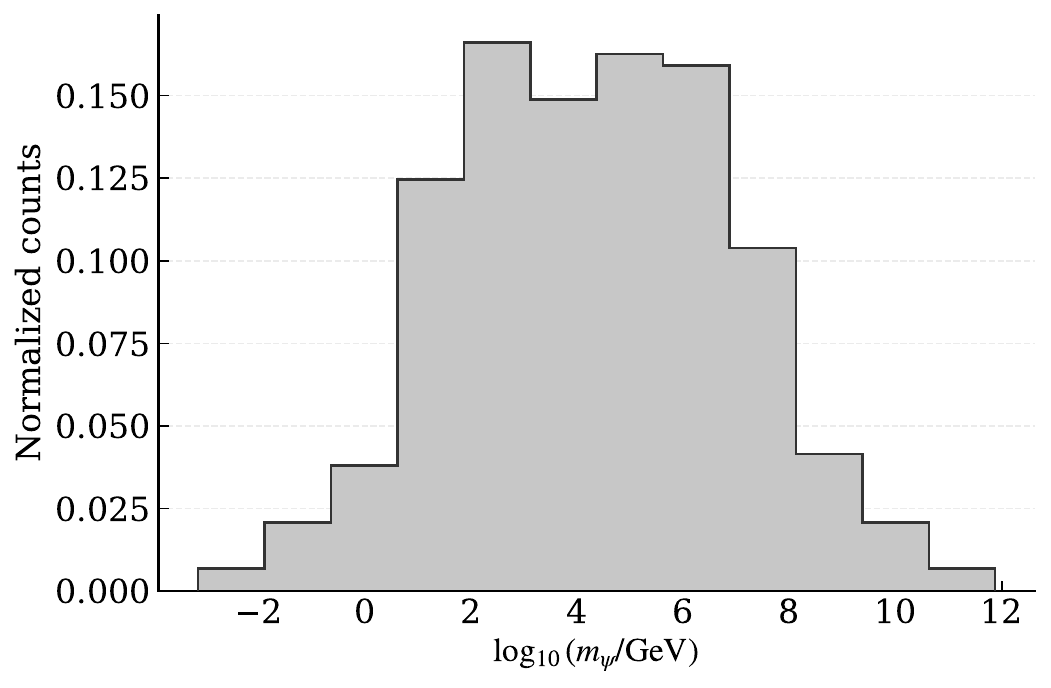}
    \caption{Normalized distribution of the DM mass \(m_\psi\) obtained from the parameter points that survive all selection criteria. 
    The distribution spans several orders of magnitude, indicating that the mechanism considered in this work is capable of producing viable DM candidates over a broad mass range. 
    We stress, however, that this numerical result should be regarded as illustrative rather than exhaustive, since our analysis is based on a limited scan of the parameter space and therefore does not cover all viable possibilities of the framework. }
	\label{figmpsi}
\end{figure}

Figure~\ref{figmpsi} shows the distribution of the DM mass \(m_\psi\) for the parameter points that survive our final selection criteria. 
One sees that the allowed values of \(m_\psi\) are spread over several orders of magnitude, rather than being confined to a narrow mass window. 
This indicates that DM production from an inflationary FOPT, followed by the subsequent evolution considered in this work, can accommodate viable DM candidates across a wide range of masses. 
In this sense, the mechanism is not restricted to a highly tuned mass scale, but instead exhibits substantial parametric flexibility. Moreover, this broad mass range is not merely a consequence of varying phase transition parameters, but also arises from the fact that the surviving parameter points span different values of the inflationary Hubble scale \(H\). The detailed parameter values for a subset of the benchmark points are collected in Appendix~\ref{markpoint}.
 
\section{GW spectrum generated by bubble collisions at the end of inflation}
\label{sec:GW}
GWs from the early Universe have been studied in several well-motivated settings. 
In particular, inflation itself predicts a primordial tensor background, while FOPTs after reheating have long been recognized as an important source of stochastic GW from expanding and colliding bubbles \cite{Starobinsky:1979ty,Rubakov:1982df,Abbott:1984fp,Hogan:1986dsh,Witten:1984rs,Kosowsky:1992vn,Caprini:2019egz,Huber:2008hg}. 
The case considered here is closer in spirit to the latter, but differs in one essential aspect: the phase transition takes place at the end of inflation, so the GW signal is generated by bubble collisions and then propagates in an inflating background.
To estimate the GW signal generated by the FOPT during inflation, we follow Refs.~\cite{Hu:2025xdt,An:2020fff} for the inflationary propagation of a short-lived source and use the envelope-approximation result of Ref.~\cite{Huber:2008hg} as the reference spectrum generated by bubble collisions. 
Because the phase transition takes place in a vacuum-dominated inflationary background, the plasma contribution is negligible and the dominant source is the collision of highly relativistic bubble walls. 
In the standard phase transition literature, the efficiency factor \(\kappa_{\rm coll}\) parametrizes the fraction of the released vacuum energy that is transferred to the source relevant for GW production~\cite{Athron:2023xlk}. 
For a vacuum transition in the runaway wall limit, it is a good approximation to take
\begin{equation}
v_w \simeq 1,
\qquad
\kappa_{\rm coll} \simeq 1,
\end{equation}
so that neither \(v_w\) nor \(\kappa_{\rm coll}\) is treated as an independent parameter in the numerical analysis. 
The relevant input parameters then reduce to
\begin{equation}
\left\{
H_{\rm inf},\,
\frac{\beta}{H_*},\,
\frac{\rho_{\rm PT}}{\rho_{\rm tot}},\,
N_*
\right\},
\end{equation}
where \(H_\ast \simeq H_{\rm inf}\) is the Hubble scale at the time of the phase transition, \(\beta^{-1}\) characterizes the duration of the transition, \(\rho_{\rm PT}/\rho_{\rm tot}\) is the fraction of the total background energy density released by the transition, and $N_\ast$ denotes the number of e-folds between the phase transition and reheating.

\paragraph{GW spectrum used in the numerical analysis.}
To estimate the present-day GW signal, we follow Refs.~\cite{Hu:2025xdt,Huber:2008hg,An:2020fff} and use the bubble collision spectrum redshifted to today as the reference template before including the additional modification induced by the remaining inflationary expansion. 
This reference spectrum is written as
\begin{equation}
h^2 \widetilde{\Omega}_{\rm GW}(\widetilde f)
=
1.27\times 10^{-6}
\left(\frac{\rho_{\rm PT}}{\rho_{\rm inf}}\right)^2
\left(\frac{H_*}{\beta}\right)^2
\left(\frac{100}{g_*}\right)^{1/3}
\frac{(a+b)\,x^a}{a\,x^{a+b}+b},
\label{eq:GW-uninflated}
\end{equation}
where
\begin{equation}
x \equiv \frac{\widetilde f}{\widetilde f_{\rm peak}},
\qquad
(a,b)=(2.8,\,1.0).
\end{equation}
The corresponding peak frequency is parameterized as
\begin{equation}
\widetilde f_{\rm peak}
=
37.8\times 10^{6}\,{\rm Hz}\,
\left(\frac{T_{\rm Rh}}{10^{15}\,{\rm GeV}}\right)
\left(\frac{\beta}{H_*}\right)
\left(\frac{g_*}{100}\right)^{1/6},
\label{eq:GW-uninflated-fpeak}
\end{equation}
with \(T_{\rm Rh}\) the reheating temperature and \(g_*\) the effective number of relativistic degrees of freedom after reheating.

The remaining inflationary expansion modifies this template in two ways: it shifts the characteristic frequency scale and deforms the spectral shape. 
Accordingly, we write the present-day spectrum as~\cite{Hu:2025xdt}
\begin{equation}
h^2\Omega_{\rm GW}(f)
=
h^2\widetilde{\Omega}_{\rm GW}\!\left(f\,e^{N_*}\right)\,
\bar{\mathcal S}(f),
\label{eq:GW-master}
\end{equation}
where \(e^{N_*}\) accounts for the frequency redshift between production and reheating, while \(\bar{\mathcal S}(f)\) encodes the inflation-induced deformation of the spectrum.

For an idealized instantaneous source, the deformation factor is taken to be~\cite{Hu:2025xdt}
\begin{equation}
{\mathcal S}_{\rm inst}(f)
=
S_0(f)+S_1(f),
\label{eq:Sinst}
\end{equation}
with
\begin{equation}
S_0(f)
=
\left[
\frac{\cos\!\bigl(\omega (1-\zeta)\bigr)}{\omega^2}
-
\frac{\sin\!\bigl(\omega (1-\zeta)\bigr)}{\omega^3}
\right]^2,
\label{eq:S0}
\end{equation}
and
\begin{align}
S_1(f)
&=
\omega\zeta
\Bigg[
\left(
\frac{1}{\omega^2}+2\zeta-1
\right)
\frac{\sin\!\bigl(2\omega(1-\zeta)\bigr)}{\omega^4}
-
\left(
\frac{2-\zeta}{\omega^2}+\zeta
\right)
\frac{\cos\!\bigl(2\omega(1-\zeta)\bigr)}{\omega^3}
\nonumber\\
&\hspace{3.3cm}
+
\frac{\zeta^3}{\omega}
\left(
\frac{1}{\omega^2}+1
\right)
\Bigg].
\label{eq:S1}
\end{align}
Here
\begin{equation}
\zeta \equiv \frac{a_*}{a_{\rm Rh}} = e^{-N_*},
\qquad
\omega \equiv \frac{2\pi a_0 f}{a_* H_*}
      = \frac{2\pi f\,e^{N_*+N_{\rm today}}}{H_*}.
\label{eq:ydef}
\end{equation}
The function \({\mathcal S}_{\rm inst}(f)\) captures the frequency-dependent distortion generated by the propagation of tensor modes during the remaining inflationary stage and by the matching to the post-inflationary evolution. 
In the infrared limit,
\begin{equation}
{\mathcal S}_{\rm inst}(f)\to \frac{1}{9},
\end{equation}
while \(S_0(f)\) provides the leading contribution for \(\omega\zeta\ll 1\). 
The term \(S_1(f)\) is subleading in the low-frequency regime, but can become important once \(\omega\zeta\gg 1\).

In our setup, however, the source is not exactly instantaneous, but has a short duration of order \(\Delta t\sim \beta^{-1}\). 
To incorporate this effect, we follow Ref.~\cite{Hu:2025xdt} and replace \({\mathcal S}_{\rm inst}(f)\) by a smeared deformation factor,
\begin{equation}
\bar{\mathcal S}(f)
=
\frac{1}{\Delta \omega}
\int_{\bar \omega-\Delta \omega/2}^{\bar \omega+\Delta \omega/2}
d\omega'\,
{\mathcal S}_{\rm inst}(\omega'),
\qquad
\bar \omega \equiv \frac{2\pi a_0 f}{a_* H_*}.
\label{eq:Sbar}
\end{equation}
This averaging prescription effectively accounts for the finite source duration in the deformation factor. 
Estimating the conformal-time duration as
\begin{equation}
\Delta\tau \sim \frac{1}{a_*\beta},
\end{equation}
one finds
\begin{equation}
\Delta \omega
\sim
a_* H_* \Delta\tau\, \bar \omega
\simeq
\frac{H_*}{\beta}\,\bar \omega.
\label{eq:Dy}
\end{equation}
Hence, in the limit \(\beta/H_* \gg 1\), the smearing effect becomes negligible, so that \(\Delta \omega \to 0\) and \(\bar{\mathcal S}(f)\to {\mathcal S}_{\rm inst}(f)\).

\paragraph{Reheating scale and redshift to today.}
To determine the peak frequency today, we further assume instantaneous reheating and identify
\begin{equation}
H_{\rm Rh} \simeq H_* \simeq H_{\rm inf}.
\end{equation}
The reheating temperature is then estimated from
\begin{equation}
3  M_{\rm Pl}^2 H_{\rm Rh}^2
\simeq
\frac{\pi^2}{30}\, g_*\, T_{\rm Rh}^4,
\end{equation}
which gives
\begin{equation}
T_{\rm Rh}
\simeq
\left(
\frac{90}{\pi^2 g_*}
\right)^{1/4}
\sqrt{ M_{\rm Pl} H_{\rm inf}},
\label{eq:TRh}
\end{equation}
The subsequent redshift from reheating to today is encoded in \(N_{\rm today}\), which in our numerical treatment is approximated as
\begin{equation}
N_{\rm today}
\simeq
\ln\!\left(\frac{T_{\rm Rh}}{T_0}\right),
\label{eq:N0}
\end{equation}
where \(T_0=2.7255\,{\rm K}\) is the CMB temperature today~\cite{Fixsen:2009ug,ParticleDataGroup:2024cfk}. 
In this approximation, we neglect the change in the effective entropy degrees of freedom during the cosmic expansion, since this effect is not important for the present purpose.

Combining Eqs.~\eqref{eq:GW-uninflated}--\eqref{eq:N0}, we obtain the GW spectrum used in our numerical analysis. 
A noteworthy effect of the remaining inflationary expansion is that the peak position of the final spectrum is not fixed by a simple redshift of the reference bubble-collision peak alone. 
Instead, it emerges from the combined effect of the redshifted collision template and the frequency dependent deformation function \(\bar{\mathcal S}(f)\)~\cite{Hu:2025xdt,An:2020fff,An:2022cce,An:2023jxf}.

\begin{figure}[h]
	\centering
	\includegraphics[width=0.8\textwidth]{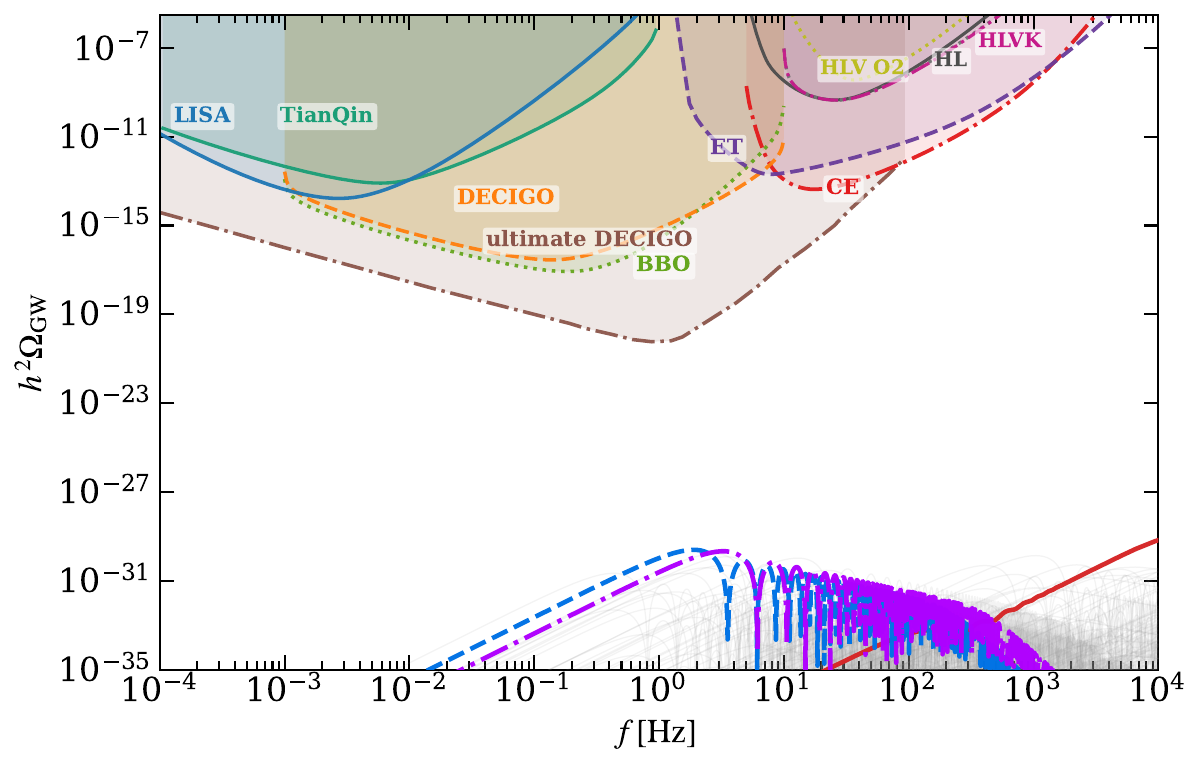}
    \caption{Present-day GW spectra for the three benchmark points with the largest values of \(h^2\Omega_{\rm GW}\) within the plotted frequency range, together with the projected experimental sensitivities. All sensitivity curves are shown for \({\rm SNR}=1\). The TianQin sensitivity is taken from Ref.~\cite{Luo:2025ewp}, the ultimate DECIGO sensitivity from Ref.~\cite{Ringwald:2020vei}, and the remaining detector sensitivities from Ref.~\cite{Schmitz:2020syl}. The red solid, blue dashed, and purple dash-dotted curves correspond respectively to the three benchmark points
    \((\mu,\epsilon,\kappa,c_{\rm{inf}},H,\beta/H)=(2.05\times10^{8}\,{\rm GeV},\,5.31\times10^{8}\,{\rm GeV},\,1.8,\,3.59\times10^{-10},\,1.18\times10^{1}\,{\rm GeV},\,6.00\times10^{2})\),
    \((1.12\times10^{11}\,{\rm GeV},\,2.36\times10^{11}\,{\rm GeV},\,1.8,\,5.99\times10^{-8},\,2.28\times10^{5}\,{\rm GeV},\,7.01\times10^{2})\), and
    \((1.52\times10^{12}\,{\rm GeV},\,2.96\times10^{12}\,{\rm GeV},\,1.8,\,7.74\times10^{-7},\,3.16\times10^{7}\,{\rm GeV},\,7.93\times10^{2})\).
    The gray line segments denote the remaining parameter points in the sample.}
	\label{figGW}
\end{figure}

Figure~\ref{figGW} presents the present-day GW spectra for the viable parameter points obtained in our scan. 
The gray line segments denote the remaining parameter points, while the red solid, blue dashed, and purple dash-dotted curves highlight the three benchmark points with the largest values of \(h^2\Omega_{\rm GW}\) in the plotted frequency range. 
As can be seen from Eq.~\eqref{eq:GW-uninflated}, the GW energy spectrum from bubble collisions is proportional to the square of the released vacuum-energy fraction, \(\Omega_{\rm GW}\propto (\rho_{\rm PT}/\rho_{\rm tot})^2\). 
In our setup, however, the spectator sector is required to remain subdominant throughout the inflationary evolution. 
In particular, as a benchmark requirement for keeping the backreaction of the phase-transition sector on the inflationary background under control, we impose that the vacuum-energy variation along the continuously occupied branch remain below \(0.1\,\rho_{\rm inf}\), as given in Eq.~\eqref{occupequ}.
Since during inflation the total energy density is dominated by the inflaton sector, \(\rho_{\rm tot}\simeq \rho_{\rm inf}\), this condition already implies that the relevant energy fraction \(\rho_{\rm PT}/\rho_{\rm tot}\) is parametrically small. 
Consequently, even before taking into account the additional suppression from the factor \((H_*/\beta)^2\) and the subsequent inflationary deformation of the spectrum, the GW amplitude is expected to be significantly reduced. 
From this perspective, the fact that the resulting spectra in our benchmark region remain below the projected experimental sensitivities is not unexpected.

\section{Summary and Discussion}
\label{sec:conclusion}

In this work, we showed that a FOPT during inflation can reproduce the observed DM abundance, but only in a restricted region of parameter space. 
This viable region is selected jointly by the de~Sitter vacuum decay structure, the requirement that the transition take place sufficiently late while still completing during inflation, and the nonequilibrium production and subsequent dilution of particles generated by runaway bubble collisions.

It has long been recognized that vacuum decay during inflation differs qualitatively from the familiar flat-space case.
Once gravitational effects are included, the CDL and Hawking--Moss channels compete, the CDL branch can disappear in part of parameter space, and additional negative modes may arise. 
As a result, not every bounce solution corresponds to a physically relevant cosmological transition. 
After imposing the conditions discussed in the main text, including the requirement that the CDL channel remains the relevant tunneling branch and that the nucleated bubbles are physically meaningful in an inflating background, the allowed parameter space is reduced to a restricted subset. 
In particular, the viable transition is not a generic consequence of a metastable potential, but instead requires a carefully selected region in which the de Sitter tunneling solution remains both mathematically well defined and cosmologically viable.

As established in previous studies, the phase transition can occur only within a restricted time window during inflation.
To avoid the appearance of excessively large bubbles and the premature disruption of the inflationary stage, the nucleation rate must remain strongly suppressed during the earlier part of inflation. 
On the other hand, for the transition to complete efficiently, the nucleation rate must increase rapidly near the end of inflation and reach the regime relevant for percolation and bubble collisions. 
This means that the viable scenario is intrinsically a late-time inflationary phase transition. 
The surviving parameter points therefore correspond to a highly time-dependent tunneling history: negligible nucleation at early times, followed by a rapid enhancement of the decay probability close to the end of inflation, together with a sufficiently large $\beta/H$ to ensure a sufficiently sharp transition.

Within this viable region, we then analyzed the particle production associated with runaway bubble collisions and the subsequent post-collision evolution. 
The final DM abundance receives contributions from both direct production in the collision process and indirect production from the later decay of the spectator field. 
In the simplified regime selected by our final parameter scan, bubble collisions effectively populate three kinds of excitations: the inflaton fluctuation, the spectator field particles, and the dark fermion sector. In the viable region, elastic self-scatterings in the spectator sector are efficient enough to redistribute the momenta of the produced \(h\) particles, and the decay \(h\to\psi\bar\psi\) dominantly converts the spectator population into dark matter. By contrast, the other scattering and number-changing processes considered in our analysis remain inefficient throughout the surviving parameter region, with reaction rates satisfying $\Gamma/H \ll 1$, and therefore do not significantly affect the final yield. 
Consequently, the present-day relic abundance is controlled, to a good approximation, by the redshifted sum of the directly produced dark fermions and the DM generated from the decay of the post-collision spectator population. 
We stress, however, that our treatment of the post-collision particle phase-space distributions is necessarily approximate, so the resulting relic-density estimate should be regarded as accurate only at the order-of-magnitude level. 
In particular, our numerical results typically show that the number density of spectator particles produced in bubble collisions is about one order of magnitude larger than that of the directly produced dark fermions. 
Therefore, even if the estimated values of $\Gamma/H$ carry non-negligible uncertainties, so that some of the scattering processes become partially competitive with the decay channel and reduce the actual amount of DM produced from spectator decay, the DM mass inferred from Eq.~\eqref{eq:rhoDMtoday_match} would still shift only within the same order of magnitude. 
For this reason, such uncertainties are not expected to qualitatively alter our main conclusions.

An additional observable consequence of this scenario is a stochastic GW background generated by bubble expansion and collision during the phase transition. 
As emphasized in studies of FOPTs during inflation, the subsequent inflationary expansion does not merely redshift the signal to lower frequencies, but can also distort its spectral shape and imprint characteristic oscillatory features~\cite{An:2020fff,Hu:2025xdt}. 
The present-day amplitude and peak frequency therefore depend not only on the microscopic parameters of the phase transition, such as the released vacuum energy and the inverse duration parameter, but also sensitively on how many e-folds of inflation remain after the bubbles collide. 
The GW signal thus provides a complementary probe of both the phase transition itself and the subsequent inflationary history. 
In the present setup, and within the assumptions adopted in our analysis, the predicted signal remains below the projected sensitivities of future GW experiments. We emphasize, however, that this conclusion is based on a numerical scan over only a limited region of parameter space, and should therefore be understood as applying only to the class of scenarios explored in this work rather than as a general statement about the entire model parameter space.

There are several natural directions for future work. 
First, it would be valuable to incorporate gravitational backreaction on the bounce more completely and to go beyond the approximations adopted here in the treatment of the inflationary background. 
Second, the present treatment of particle production from bubble collisions is still rather simplified. 
A more complete analysis would require a better-motivated treatment of the momentum integration range entering the production estimates, a more microscopic determination of the distribution functions of the produced particles, and a more careful study of their subsequent nonequilibrium evolution. 
In addition, recent work has shown that particles produced in bubble collisions can themselves source an additional stochastic GW component, which has not been included in the present analysis and deserves further investigation~\cite{Inomata:2024rkt}. 
Finally, it would be interesting to embed this mechanism into more specific
particle-physics realizations of the dark sector and to explore additional
observational signatures beyond the primary GW signal. These may include
reheating-dependent modifications of the GW spectrum~\cite{An:2020fff,An:2022cce,Barir:2022kzo,Garcia:2026hfd}, as well as possible dark-radiation isocurvature signatures in scenarios where the
phase transition products behave as dark radiation~\cite{Buckley:2024nen}. More generally, a vacuum transition in a subdominant non-inflaton sector during
inflation can also source imprints in the primordial curvature perturbation,
including possible non-Gaussian signatures, if the transition rate or transition
time is modulated by fluctuations of an additional light field~\cite{Garbrecht:2025crs}. 
Other cosmological consequences, such as baryogenesis-related effects, could
also arise if the dark sector communicates with the visible sector~\cite{Bhusal:2025lvm}.

\paragraph{Acknowledgements:} 
We would like to thank Aidi Yang for helpful discussions. This work is supported by the National Natural Science Foundation of China (NNSFC) Grant
No.~12475111, No.~12205387, and the Fundamental
Research Funds for the Central Universities, Sun Yat-sen University.

\newpage

\appendix

\section{Cross Sections Relevant for DM Production}
\label{app:cross_sections}

\subsection{Organization of the relevant subprocesses}

The post-collision dynamics relevant for our analysis are governed by the interaction terms
\begin{equation}
\mathcal{L}
\supset
-\frac{\lambda\epsilon}{3}h^3
-\frac{\kappa}{4}h^4
-y\,h\,\bar{\psi}\psi
+c_{\rm inf}^2\bar{\phi}\,\varphi_{\rm{inf}}\,h^2
+\frac{c_{\rm inf}^2}{2}\varphi_{\rm{inf}}^2 h^2,
\end{equation}
where the first two terms describe the cubic and quartic self-interactions of the spectator field \(h\), the Yukawa interaction \(y\,h\,\bar\psi\psi\) allows \(h\) to decay into DM particles, and the last two terms encode the couplings between \(h\) and the inflaton background \(\bar\phi\) as well as the inflaton fluctuation \(\varphi_{\rm{inf}}\). 
Based on these interactions, and restricting attention to the channels most directly relevant to the final DM abundance, we include only the following six subprocesses in the post-collision evolution:
\begin{enumerate}
    \item \emph{Direct DM production}:
    \begin{equation}
        h \to \psi \bar\psi,
        \qquad
        h h \to \psi \bar\psi .
    \end{equation}
    These are the channels that directly convert the spectator field into DM.  The \(\psi\) particles produced during the bubble-collision stage are expected to carry a characteristic energy comparable to that of the simultaneously produced \(h\) excitations. 
    In the viable parameter region obtained in our analysis, the characteristic energies of the two species are consistently found to be of the same order. Therefore, even if the inverse decay process \(\psi\bar\psi \to h\) occurs, it would only repopulate \(h\) at a comparable characteristic energy and would not appreciably modify the typical energy of the \(h\) population; the regenerated \(h\) particles subsequently decay again through \(h\to\psi\bar\psi\). For this reason, we neglect the inverse process in the following.

    \item \emph{Scalar self-interaction channels}:
    \begin{equation}
        h h \to h h,
        \qquad
        3h \to 2h .
    \end{equation}
    These processes do not directly produce DM, but they can affect the scalar abundance before the decay \(h\to\psi\bar\psi\) is completed. 
    In particular, the elastic channel \(hh\to hh\) controls momentum redistribution inside the scalar sector, while the number-changing process \(3h\to2h\) can reduce the number of \(h\) particles and thereby suppress the final DM abundance relative to the naive expectation based on decay alone. 
    For the parameter region of interest, however, the typical energy of the produced \(h\) particles is only of order \(m_h\), so the inverse process \(2h\to3h\) already lies close to kinematic threshold and is therefore strongly suppressed. Even if it occurs occasionally, it does not modify the \(h\) abundance at the order-of-magnitude level, and we therefore neglect it in the following.

    \item \emph{Competing sink channels involving inflaton fluctuations}:
    \begin{equation}
        h h \to \varphi_{\rm{inf}}\varphi_{\rm{inf}},
        \qquad
        h h \to h\varphi_{\rm{inf}} .
    \end{equation}
    In the present work, we focus on the regime in which the energy stored in the phase-transition field is ultimately transferred into DM. 
    From this perspective, channels involving the inflaton fluctuation \(\varphi_{\rm{inf}}\) are treated only as possible competing sink channels that may reduce the fraction of the scalar sector converted into DM. 
    Moreover, in the parameter region of interest, our results show that the abundance of \(\varphi_{\rm{inf}}\) particles produced during bubble collisions is much smaller than that of both \(h\) and \(\psi\); accordingly, the corresponding inverse processes are negligible in practice and will be safely neglected in the following.
\end{enumerate}

\subsection{Cross sections and decay width used in the rate estimates}
\label{app:cross_sections_used}

\subsubsection{\texorpdfstring{$h\to\psi\bar\psi$}{h -> psi psibar}}
\begin{figure}[h]
\centering
\begin{tikzpicture}
\begin{feynman}
    \vertex (a) at (0,0);
    \vertex [left=1.6cm of a] (i1) {$h$};
    \vertex [right=1.6cm of a] (f1) {$\psi$};
    \vertex [below right=1.2cm and 1.6cm of a] (f2) {$\bar{\psi}$};

    \diagram* {
        (i1) -- [scalar] (a),
        (a) -- [fermion] (f1),
        (a) -- [anti fermion] (f2),
    };
\end{feynman}
\end{tikzpicture}
\caption{Feynman diagram for the decay process $h\to \psi\bar{\psi}$.}
\label{fig:h_to_psipsibar}
\end{figure}
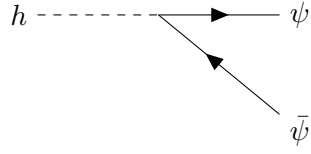

As shown in Fig.~\ref{fig:h_to_psipsibar}, if the decay is kinematically open, $m_h>2m_\psi$, the tree-level width is
\begin{equation}
\Gamma_{h\to\psi\bar\psi}
=
\frac{y^2\,m_h}{8\pi}
\left(1-\frac{4m_\psi^2}{m_h^2}\right)^{3/2}.
\label{eq:app_decay_width}
\end{equation}
This is the reference channel against which we compare the depletion effects of scalar self-interactions.

\subsubsection{\texorpdfstring{$hh\to\psi\bar\psi$}{hh -> psi psibar}}
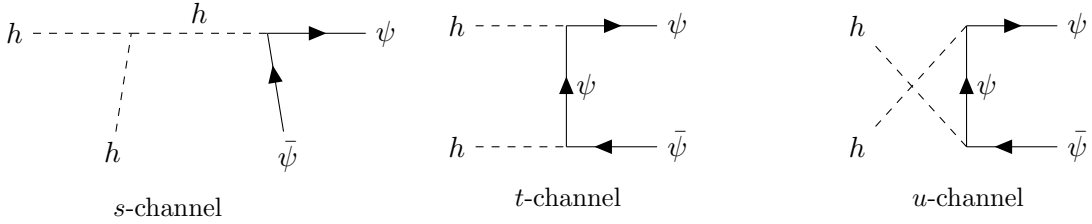
\begin{figure}[h]
\centering

\begin{minipage}{0.3\textwidth}
\centering
\begin{tikzpicture}
\begin{feynman}
    \vertex (v1) at (0,0);
    \vertex (v2) at (1.8,0);
    \vertex [left=1.3cm of v1] (i1) {$h$};
    \vertex [below left=1.3cm and 0cm of v1] (i2) {$h$};
    \vertex [right=1.3cm of v2] (f1) {$\psi$};
    \vertex [below right=1.3cm and 0cm of v2] (f2) {$\bar{\psi}$};

    \diagram* {
        (i1) -- [scalar] (v1),
        (i2) -- [scalar] (v1),
        (v1) -- [scalar, edge label=$h$] (v2),
        (v2) -- [fermion] (f1),
        (v2) -- [anti fermion] (f2),
    };
\end{feynman}
\end{tikzpicture}

\small $s$-channel
\end{minipage}
\hfill
\begin{minipage}{0.3\textwidth}
\centering
\begin{tikzpicture}
\begin{feynman}
    \vertex (vu) at (0,0.8);
    \vertex (vd) at (0,-0.8);
    \vertex [left=1.2cm of vu] (i1) {$h$};
    \vertex [left=1.2cm of vd] (i2) {$h$};
    \vertex [right=1.2cm of vu] (f1) {$\psi$};
    \vertex [right=1.2cm of vd] (f2) {$\bar{\psi}$};

    \diagram* {
        (i1) -- [scalar] (vu),
        (i2) -- [scalar] (vd),
        (vd) -- [fermion, edge label'=$\psi$] (vu),
        (vu) -- [fermion] (f1),
        (vd) -- [anti fermion] (f2),
    };
\end{feynman}
\end{tikzpicture}

\small $t$-channel
\end{minipage}
\hfill
\begin{minipage}{0.3\textwidth}
\centering
\begin{tikzpicture}
\begin{feynman}
    \vertex (vu) at (0,0.8);
    \vertex (vd) at (0,-0.8);
    \vertex [left=1.2cm of vu] (i1) {$h$};
    \vertex [left=1.2cm of vd] (i2) {$h$};
    \vertex [right=1.2cm of vu] (f1) {$\psi$};
    \vertex [right=1.2cm of vd] (f2) {$\bar{\psi}$};

    \diagram* {
        (i1) -- [scalar] (vd),
        (i2) -- [scalar] (vu),
        (vd) -- [fermion, edge label'=$\psi$] (vu),
        (vu) -- [fermion] (f1),
        (vd) -- [anti fermion] (f2),
    };
\end{feynman}
\end{tikzpicture}

\small $u$-channel
\end{minipage}

\caption{Tree-level Feynman diagrams for the process $hh\to \psi\bar{\psi}$.}
\label{fig:hh_to_psipsibar}
\end{figure}

As shown in Fig.~\ref{fig:hh_to_psipsibar}, the direct production channel is
\begin{equation}
h h \to \psi\bar\psi.
\end{equation}
For convenience, in the formulas below we define
\begin{equation}
M \equiv m_h,
\qquad
m \equiv m_\psi ,
\end{equation}
the total cross section is
\begin{align}
\sigma(hh\to\psi\bar\psi)
&=
\frac{\beta_\psi}{32\pi s\,\beta_h}
\Bigg[
\frac{16(\lambda\epsilon)^2y^2(s-4m^2)}{(s-M^2)^2}
-\frac{4y^4\Big(M^4+(M^2-4m^2)^2\Big)}{B^2-k^2}
\nonumber\\
&\hspace{4em}
+\frac{L(s)}{k}
\left(
\frac{16\lambda\epsilon m y^3 (s-4m^2)}{s-M^2}
-16m^2y^4
-\frac{4y^4\big(s^2-4sM^2+2M^4\big)}{s-2M^2}
\right)
\Bigg],
\label{eq:app_sigma_hh_to_psipsibar_final}
\end{align}
where
\begin{equation}
\beta_h \equiv \sqrt{1-\frac{4M^2}{s}},
\qquad
\beta_\psi \equiv \sqrt{1-\frac{4m^2}{s}},
\qquad
k \equiv \frac{s}{2}\beta_h\beta_\psi,
\qquad
B \equiv M^2-\frac{s}{2},
\end{equation}
and
\begin{equation}
L(s)\equiv \ln\left(\frac{B+k}{B-k}\right).
\end{equation}
This is the main \(2\to2\) channel that directly converts the spectator particles into DM when it becomes efficient.

\subsubsection{\texorpdfstring{$hh\to hh$}{hh -> hh}}
\begin{figure}[h]
\centering
\begin{minipage}{0.24\textwidth}
\centering
\begin{tikzpicture}
\begin{feynman}
    \vertex (v) at (0,0);
    \vertex [above left=1.1cm and 1.1cm of v] (i1) {$h$};
    \vertex [below left=1.1cm and 1.1cm of v] (i2) {$h$};
    \vertex [above right=1.1cm and 1.1cm of v] (f1) {$h$};
    \vertex [below right=1.1cm and 1.1cm of v] (f2) {$h$};

    \diagram* {
        (i1) -- [scalar] (v) -- [scalar] (f1),
        (i2) -- [scalar] (v) -- [scalar] (f2),
    };
\end{feynman}
\end{tikzpicture}

\small contact
\end{minipage}
\hfill
\begin{minipage}{0.24\textwidth}
\centering
\begin{tikzpicture}
\begin{feynman}
    \vertex (v1) at (0,0);
    \vertex (v2) at (1.8,0);
    \vertex [above left=0.9cm and 1.0cm of v1] (i1) {$h$};
    \vertex [below left=0.9cm and 1.0cm of v1] (i2) {$h$};
    \vertex [above right=0.9cm and 1.0cm of v2] (f1) {$h$};
    \vertex [below right=0.9cm and 1.0cm of v2] (f2) {$h$};

    \diagram* {
        (i1) -- [scalar] (v1),
        (i2) -- [scalar] (v1),
        (v1) -- [scalar, edge label=$h$] (v2),
        (v2) -- [scalar] (f1),
        (v2) -- [scalar] (f2),
    };
\end{feynman}
\end{tikzpicture}

\small $s$-channel ($h$)
\end{minipage}
\hfill
\begin{minipage}{0.24\textwidth}
\centering
\begin{tikzpicture}
\begin{feynman}
    \vertex (vu) at (0,0.85);
    \vertex (vd) at (0,-0.85);
    \vertex [left=1.2cm of vu] (i1) {$h$};
    \vertex [left=1.2cm of vd] (i2) {$h$};
    \vertex [right=1.2cm of vu] (f1) {$h$};
    \vertex [right=1.2cm of vd] (f2) {$h$};

    \diagram* {
        (i1) -- [scalar] (vu),
        (i2) -- [scalar] (vd),
        (vd) -- [scalar, edge label'=$h$] (vu),
        (vu) -- [scalar] (f1),
        (vd) -- [scalar] (f2),
    };
\end{feynman}
\end{tikzpicture}

\small $t$-channel ($h$)
\end{minipage}
\hfill
\begin{minipage}{0.24\textwidth}
\centering
\begin{tikzpicture}
\begin{feynman}
    \vertex (vu) at (0,0.85);
    \vertex (vd) at (0,-0.85);
    \vertex [left=1.2cm of vu] (i1) {$h$};
    \vertex [left=1.2cm of vd] (i2) {$h$};
    \vertex [right=1.2cm of vu] (f1) {$h$};
    \vertex [right=1.2cm of vd] (f2) {$h$};

    \diagram* {
        (i1) -- [scalar] (vd),
        (i2) -- [scalar] (vu),
        (vd) -- [scalar, edge label'=$h$] (vu),
        (vu) -- [scalar] (f1),
        (vd) -- [scalar] (f2),
    };
\end{feynman}
\end{tikzpicture}

\small $u$-channel ($h$)
\end{minipage}

\vspace{0.8cm}

\begin{minipage}{0.24\textwidth}
\centering
\begin{tikzpicture}
\begin{feynman}
    \vertex (v1) at (0,0);
    \vertex (v2) at (1.8,0);
    \vertex [above left=0.9cm and 1.0cm of v1] (i1) {$h$};
    \vertex [below left=0.9cm and 1.0cm of v1] (i2) {$h$};
    \vertex [above right=0.9cm and 1.0cm of v2] (f1) {$h$};
    \vertex [below right=0.9cm and 1.0cm of v2] (f2) {$h$};

    \diagram* {
        (i1) -- [scalar] (v1),
        (i2) -- [scalar] (v1),
        (v1) -- [scalar, edge label=$\varphi_{\rm{inf}}$] (v2),
        (v2) -- [scalar] (f1),
        (v2) -- [scalar] (f2),
    };
\end{feynman}
\end{tikzpicture}

\small $s$-channel ($\varphi_{\rm{inf}}$)
\end{minipage}
\hfill
\begin{minipage}{0.24\textwidth}
\centering
\begin{tikzpicture}
\begin{feynman}
    \vertex (vu) at (0,0.85);
    \vertex (vd) at (0,-0.85);
    \vertex [left=1.2cm of vu] (i1) {$h$};
    \vertex [left=1.2cm of vd] (i2) {$h$};
    \vertex [right=1.2cm of vu] (f1) {$h$};
    \vertex [right=1.2cm of vd] (f2) {$h$};

    \diagram* {
        (i1) -- [scalar] (vu),
        (i2) -- [scalar] (vd),
        (vd) -- [scalar, edge label'=$\varphi_{\rm{inf}}$] (vu),
        (vu) -- [scalar] (f1),
        (vd) -- [scalar] (f2),
    };
\end{feynman}
\end{tikzpicture}

\small $t$-channel ($\varphi_{\rm{inf}}$)
\end{minipage}
\hfill
\begin{minipage}{0.24\textwidth}
\centering
\begin{tikzpicture}
\begin{feynman}
    \vertex (vu) at (0,0.85);
    \vertex (vd) at (0,-0.85);
    \vertex [left=1.2cm of vu] (i1) {$h$};
    \vertex [left=1.2cm of vd] (i2) {$h$};
    \vertex [right=1.2cm of vu] (f1) {$h$};
    \vertex [right=1.2cm of vd] (f2) {$h$};

    \diagram* {
        (i1) -- [scalar] (vd),
        (i2) -- [scalar] (vu),
        (vd) -- [scalar, edge label'=$\varphi_{\rm{inf}}$] (vu),
        (vu) -- [scalar] (f1),
        (vd) -- [scalar] (f2),
    };
\end{feynman}
\end{tikzpicture}

\small $u$-channel ($\varphi_{\rm{inf}}$)
\end{minipage}

\caption{Tree-level Feynman diagrams for the process $hh\to hh$. The first row shows the contact interaction and the $h$-exchange diagrams, while the second row shows the $\varphi_{\rm{inf}}$-exchange diagrams.}
\label{fig:hh_to_hh}
\end{figure}
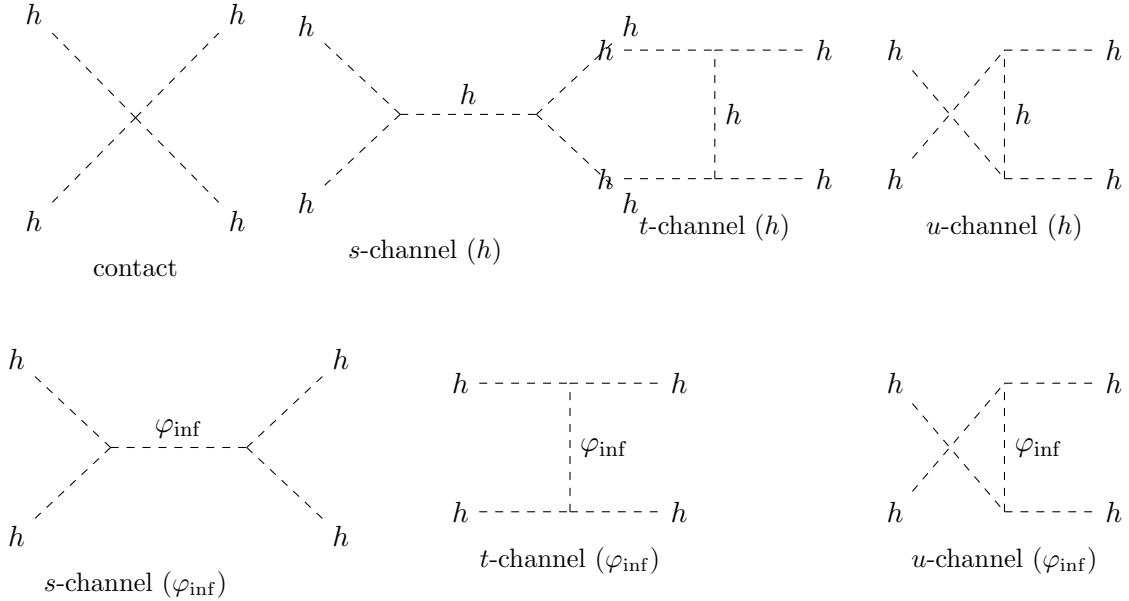

As shown in Fig.~\ref{fig:hh_to_hh}, for the elastic scalar self-scattering channel
\begin{equation}
hh\to hh,
\end{equation}
and assuming $m_{\varphi_{\rm{inf}}}=m_h$ so that the $h$- and $\varphi_{\rm{inf}}$-exchange contributions can be combined, the total cross section is
\begin{equation}
\sigma(hh\to hh)
=
\frac{1}{64\pi s}
\left[
2C_0^2+2C_0K\,I_1(a,b)+K^2 I_2(a,b)
\right],
\label{eq:app_sigma_hhhh_final}
\end{equation}
where
\begin{equation}
G^2\equiv (\lambda\epsilon)^2+c_{\rm inf}^4\bar\phi^{\,2},
\qquad
a\equiv \frac{s}{2}-m_h^2,
\qquad
b\equiv \frac{s}{2}-2m_h^2,
\end{equation}
\begin{equation}
C_0=-6\kappa-\frac{4G^2}{s-m_h^2},
\qquad
K=8aG^2,
\end{equation}
and
\begin{equation}
I_1(a,b)=\frac{1}{ab}\ln\!\left(\frac{a+b}{a-b}\right),
\end{equation}
\begin{equation}
I_2(a,b)=
\frac{2ab+(a^2-b^2)\ln\!\left(\frac{a+b}{a-b}\right)}
{2a^3b(a^2-b^2)}.
\end{equation}
This channel does not change the total scalar number, but it controls the internal redistribution of momentum in the scalar sector and is therefore useful for estimating the characteristic spectator energy entering the effective rate calculation.

\subsubsection{\texorpdfstring{$hh\to\varphi_{\rm{inf}}\varphi_{\rm{inf}}$}{hh -> varphi varphi}}
\begin{figure}[h]
\centering

\begin{minipage}{0.3\textwidth}
\centering
\begin{tikzpicture}
\begin{feynman}
    \vertex (v) at (0,0);
    \vertex [above left=1.1cm and 1.1cm of v] (i1) {$h$};
    \vertex [below left=1.1cm and 1.1cm of v] (i2) {$h$};
    \vertex [above right=1.1cm and 1.1cm of v] (f1) {$\varphi_{\rm{inf}}$};
    \vertex [below right=1.1cm and 1.1cm of v] (f2) {$\varphi_{\rm{inf}}$};

    \diagram* {
        (i1) -- [scalar] (v) -- [scalar] (f1),
        (i2) -- [scalar] (v) -- [scalar] (f2),
    };
\end{feynman}
\end{tikzpicture}

\small contact
\end{minipage}
\hfill
\begin{minipage}{0.3\textwidth}
\centering
\begin{tikzpicture}
\begin{feynman}
    \vertex (vu) at (0,0.85);
    \vertex (vd) at (0,-0.85);
    \vertex [left=1.2cm of vu] (i1) {$h$};
    \vertex [left=1.2cm of vd] (i2) {$h$};
    \vertex [right=1.2cm of vu] (f1) {$\varphi_{\rm{inf}}$};
    \vertex [right=1.2cm of vd] (f2) {$\varphi_{\rm{inf}}$};

    \diagram* {
        (i1) -- [scalar] (vu),
        (i2) -- [scalar] (vd),
        (vd) -- [scalar, edge label'=$h$] (vu),
        (vu) -- [scalar] (f1),
        (vd) -- [scalar] (f2),
    };
\end{feynman}
\end{tikzpicture}

\small $t$-channel ($h$)
\end{minipage}
\hfill
\begin{minipage}{0.3\textwidth}
\centering
\begin{tikzpicture}
\begin{feynman}
    \vertex (vu) at (0,0.85);
    \vertex (vd) at (0,-0.85);
    \vertex [left=1.2cm of vu] (i1) {$h$};
    \vertex [left=1.2cm of vd] (i2) {$h$};
    \vertex [right=1.2cm of vu] (f1) {$\varphi_{\rm{inf}}$};
    \vertex [right=1.2cm of vd] (f2) {$\varphi_{\rm{inf}}$};

    \diagram* {
        (i1) -- [scalar] (vd),
        (i2) -- [scalar] (vu),
        (vd) -- [scalar, edge label'=$h$] (vu),
        (vu) -- [scalar] (f1),
        (vd) -- [scalar] (f2),
    };
\end{feynman}
\end{tikzpicture}

\small $u$-channel ($h$)
\end{minipage}

\caption{Tree-level Feynman diagrams for the process $hh\to\varphi_{\rm{inf}}\varphi_{\rm{inf}}$. The first diagram arises from the quartic contact interaction $h^2\varphi_{\rm{inf}}^2$, while the second and third correspond to the $t$- and $u$-channel exchange of $h$ induced by the cubic vertex $\varphi_{\rm{inf}} h^2$.}
\label{fig:hh_to_varphivarphi}
\end{figure}
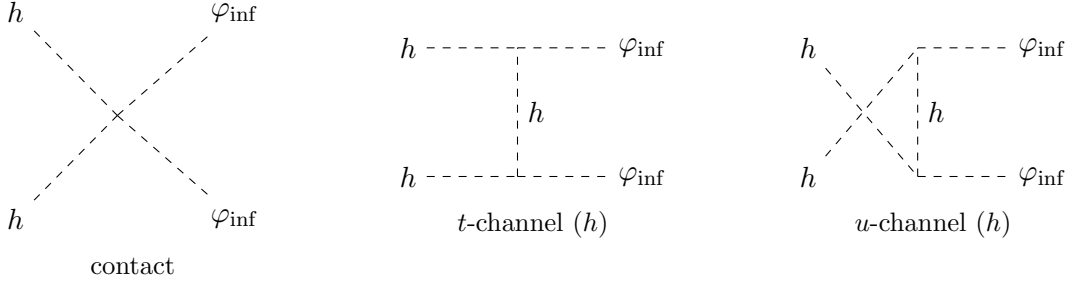

As shown in Fig.~\ref{fig:hh_to_varphivarphi}, for the competing sink channel
\begin{equation}
hh\to\varphi_{\rm{inf}}\varphi_{\rm{inf}},
\end{equation}
under the simplifying assumption \(m_{\varphi_{\rm{inf}}} = m_h\), the total cross section is
\begin{equation}
\sigma(hh\to\varphi\varphi)
=
\frac{1}{64\pi s}
\left[
2C_0^2+2C_0K\,I_1(a,b)+K^2 I_2(a,b)
\right],
\label{eq:app_sigma_hh_to_varphivarphi}
\end{equation}
where
\begin{equation}
a\equiv \frac{s}{2}-m_h^2,
\qquad
b\equiv \frac{s}{2}-2m_h^2,
\qquad
C_0=2c_{\rm inf}^2,
\qquad
K=8a\,c_{\rm inf}^4\bar\phi^{\,2},
\end{equation}
and the functions $I_1(a,b)$ and $I_2(a,b)$ are the same as above. 
If efficient, this channel reduces the fraction of the spectator particles that is eventually converted into DM.

\subsubsection{\texorpdfstring{$hh\to h\varphi_{\rm{inf}}$}{hh -> h varphi}}
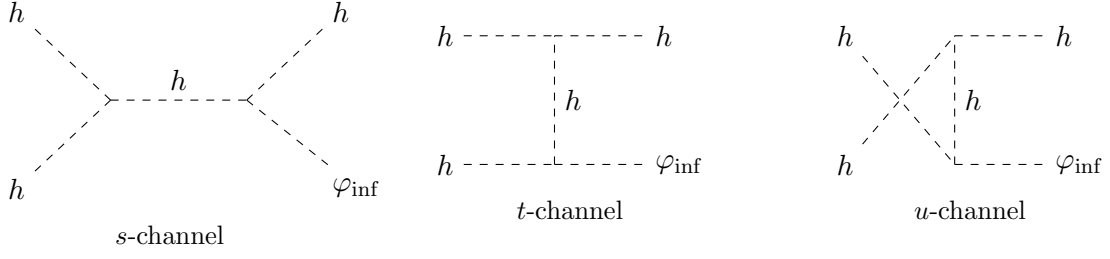
\begin{figure}[h]
\centering

\begin{minipage}{0.3\textwidth}
\centering
\begin{tikzpicture}
\begin{feynman}
    \vertex (v1) at (0,0);
    \vertex (v2) at (1.8,0);
    \vertex [above left=0.9cm and 1.0cm of v1] (i1) {$h$};
    \vertex [below left=0.9cm and 1.0cm of v1] (i2) {$h$};
    \vertex [above right=0.9cm and 1.0cm of v2] (f1) {$h$};
    \vertex [below right=0.9cm and 1.0cm of v2] (f2) {$\varphi_{\rm{inf}}$};

    \diagram* {
        (i1) -- [scalar] (v1),
        (i2) -- [scalar] (v1),
        (v1) -- [scalar, edge label=$h$] (v2),
        (v2) -- [scalar] (f1),
        (v2) -- [scalar] (f2),
    };
\end{feynman}
\end{tikzpicture}

\small $s$-channel
\end{minipage}
\hfill
\begin{minipage}{0.3\textwidth}
\centering
\begin{tikzpicture}
\begin{feynman}
    \vertex (vu) at (0,0.85);
    \vertex (vd) at (0,-0.85);
    \vertex [left=1.2cm of vu] (i1) {$h$};
    \vertex [left=1.2cm of vd] (i2) {$h$};
    \vertex [right=1.2cm of vu] (f1) {$h$};
    \vertex [right=1.2cm of vd] (f2) {$\varphi_{\rm{inf}}$};

    \diagram* {
        (i1) -- [scalar] (vu),
        (i2) -- [scalar] (vd),
        (vd) -- [scalar, edge label'=$h$] (vu),
        (vu) -- [scalar] (f1),
        (vd) -- [scalar] (f2),
    };
\end{feynman}
\end{tikzpicture}

\small $t$-channel
\end{minipage}
\hfill
\begin{minipage}{0.3\textwidth}
\centering
\begin{tikzpicture}
\begin{feynman}
    \vertex (vu) at (0,0.85);
    \vertex (vd) at (0,-0.85);
    \vertex [left=1.2cm of vu] (i1) {$h$};
    \vertex [left=1.2cm of vd] (i2) {$h$};
    \vertex [right=1.2cm of vu] (f1) {$h$};
    \vertex [right=1.2cm of vd] (f2) {$\varphi_{\rm{inf}}$};

    \diagram* {
        (i1) -- [scalar] (vd),
        (i2) -- [scalar] (vu),
        (vd) -- [scalar, edge label'=$h$] (vu),
        (vu) -- [scalar] (f1),
        (vd) -- [scalar] (f2),
    };
\end{feynman}
\end{tikzpicture}

\small $u$-channel
\end{minipage}

\caption{Tree-level Feynman diagrams for the process $hh\to h\varphi_{\rm{inf}}$.}
\label{fig:hh_to_hvarphi}
\end{figure}

As shown in Fig.~\ref{fig:hh_to_hvarphi}, for the second competing sink channel
\begin{equation}
hh\to h\varphi_{\rm{inf}},
\end{equation}
the total cross section is
\begin{equation}
\sigma(hh\to h\varphi_{\rm{inf}})
=
\frac{(\lambda\epsilon)^2g^2}{2\pi s}\,\frac{p_f}{p_i}
\left[
2S^2
+\frac{4}{a^2-b^2}
+\left(
\frac{4S}{b}+\frac{2}{ab}
\right)\ln\left|\frac{a+b}{a-b}\right|
\right],
\label{eq:app_sigma_hh_to_hvarphi}
\end{equation}
where
\begin{equation}
g\equiv c_{\rm inf}^2\bar\phi,
\qquad
S\equiv \frac{1}{s-m_h^2},
\qquad
a\equiv \frac{m_h^2+m_{\varphi_{\rm{inf}}}^2-s}{2},
\qquad
b\equiv 2p_i p_f,
\end{equation}
with
\begin{equation}
p_i=\frac{\sqrt{\lambda(s,m_h^2,m_h^2)}}{2\sqrt{s}},
\qquad
p_f=\frac{\sqrt{\lambda(s,m_h^2,m_{\varphi_{\rm{inf}}}^2)}}{2\sqrt{s}}.
\end{equation}
This process transfers part of the spectator population into final states involving $\varphi_{\rm{inf}}$ and therefore also competes with DM production.

\subsubsection{\texorpdfstring{$3h\to2h$}{3h -> 2h}}
\begin{figure}[h]
\centering

\begin{minipage}{0.46\textwidth}
\centering
\begin{tikzpicture}
\begin{feynman}
    \vertex (v1) at (0,0);
    \vertex (v2) at (2.0,0);

    \vertex [left=1.4cm of v1] (i1) {$h$};
    \vertex [above left=1.0cm and 1.0cm of v1] (i2) {$h$};
    \vertex [below left=1.0cm and 1.0cm of v1] (i3) {$h$};

    \vertex [above right=1.0cm and 1.2cm of v2] (f1) {$h$};
    \vertex [below right=1.0cm and 1.2cm of v2] (f2) {$h$};

    \diagram* {
        (i1) -- [scalar] (v1),
        (i2) -- [scalar] (v1),
        (i3) -- [scalar] (v1),
        (v1) -- [scalar, edge label=$h$] (v2),
        (v2) -- [scalar] (f1),
        (v2) -- [scalar] (f2),
    };
\end{feynman}
\end{tikzpicture}

\small $(4+3)$ topology
\end{minipage}
\hfill
\begin{minipage}{0.46\textwidth}
\centering
\begin{tikzpicture}
\begin{feynman}
    \vertex (v1) at (0,0.8);
    \vertex (v2) at (1.6,0);
    \vertex (v3) at (3.2,0.8);

    \vertex [above left=1.0cm and 1.0cm of v1] (i1) {$h$};
    \vertex [below left=1.0cm and 1.0cm of v1] (i2) {$h$};
    \vertex [below=1.4cm of v2] (i3) {$h$};

    \vertex [above right=1.0cm and 1.0cm of v3] (f1) {$h$};
    \vertex [below right=1.0cm and 1.0cm of v3] (f2) {$h$};

    \diagram* {
        (i1) -- [scalar] (v1),
        (i2) -- [scalar] (v1),
        (v1) -- [scalar, edge label=$h$] (v2),

        (i3) -- [scalar] (v2),
        (v2) -- [scalar, edge label=$h$] (v3),

        (v3) -- [scalar] (f1),
        (v3) -- [scalar] (f2),
    };
\end{feynman}
\end{tikzpicture}

\small $(3+3+3)$ topology
\end{minipage}

\caption{Representative tree-level Feynman diagrams for the process $3h\to2h$. The left panel shows the $(4+3)$ topology involving one quartic and one cubic scalar vertex, while the right panel shows the $(3+3+3)$ topology involving three cubic scalar vertices.}
\label{fig:3h_to_2h}
\end{figure}
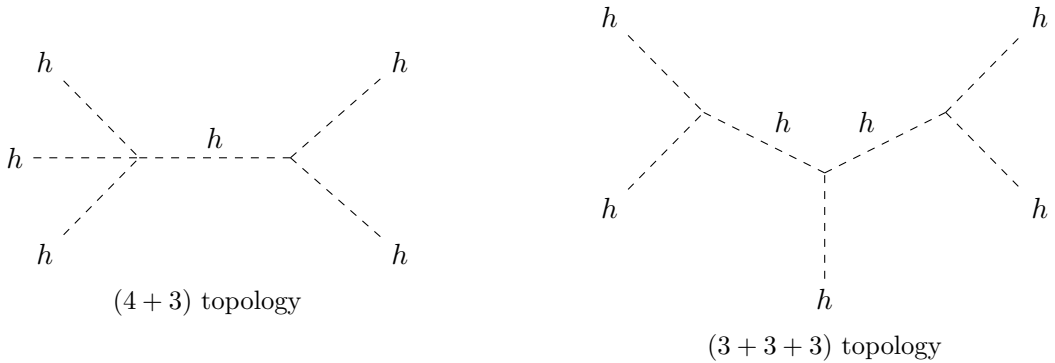

As shown in Fig.~\ref{fig:3h_to_2h}, the number-changing process
\begin{equation}
3h \to 2h
\end{equation}
is relevant because it competes with the decay channel $h\to \psi\bar\psi$. 
Our main concern is whether the $h$ particles produced after bubble collisions decay predominantly into DM, or whether a non-negligible fraction of them is first depleted by scalar self-interactions. 
If the process $3h\to 2h$ becomes efficient before the decay is completed, the comoving number density of $h$ is reduced, and the final DM abundance inherited from $h$ decay is correspondingly suppressed relative to the estimate based on decay alone.

Rather than performing a complete phase-space calculation, we employ a tree-level order-of-magnitude estimate based on naive dimensional analysis. 
It is convenient to rewrite the scalar self-interactions in the standard factorial normalization,
\begin{equation}
\mathcal L_{\rm int}
=
-\frac{\lambda_3}{3!}\,h^3
-\frac{\lambda_4}{4!}\,h^4,
\qquad
\lambda_3 = 2\lambda\epsilon,
\qquad
\lambda_4 = 6\kappa,
\end{equation}
so that the momentum-space vertex factors are simply $-i\lambda_3$ and $-i\lambda_4$.

We work in the three-body center-of-mass frame and take the incoming energies to be
\begin{equation}
E_1=E_2=E_3=N_E\,m_h.
\end{equation}
Then the total invariant mass is
\begin{equation}
\sqrt{s_{3\mathrm{body}}}=E_1+E_2+E_3=3N_E m_h,
\qquad
s_{3\mathrm{body}}=9N_E^2 m_h^2.
\end{equation}
Away from resonances and special kinematic configurations, the typical virtuality flowing through an internal scalar propagator is of order
\begin{equation}
q^2-m_h^2 \sim N_E^2 m_h^2,
\end{equation}
so that each internal propagator is estimated parametrically as
\begin{equation}
\frac{1}{q^2-m_h^2}\sim \frac{1}{N_E^2 m_h^2}.
\end{equation}

For an order-of-magnitude estimate, we retain only the two representative tree-level diagrams shown in Fig.~\ref{fig:3h_to_2h}, namely one diagram of the $(4+3)$ type and one diagram of the $(3+3+3)$ type. 
Under the above approximation, a single $(4+3)$ diagram contributes at the level
\begin{equation}
\mathcal M_{(4+3)}^{\rm single}
\sim
\frac{\lambda_3\lambda_4}{N_E^2 m_h^2},
\end{equation}
whereas a single $(3+3+3)$ diagram scales as
\begin{equation}
\mathcal M_{(3+3+3)}^{\rm single}
\sim
\frac{\lambda_3^3}{N_E^4 m_h^4}.
\end{equation}
We then estimate the full amplitude as
\begin{equation}
\mathcal M_{3\to2}^{(N_E)}
\sim
10\,\frac{\lambda_3\lambda_4}{N_E^2 m_h^2}
+
15\,\frac{\lambda_3^3}{N_E^4 m_h^4},
\end{equation}
and therefore
\begin{equation}
\left|\mathcal M_{3\to2}^{(N_E)}\right|^2
\sim
\left|
10\,\frac{\lambda_3\lambda_4}{N_E^2 m_h^2}
+
15\,\frac{\lambda_3^3}{N_E^4 m_h^4}
\right|^2.
\end{equation}

The integrated two-body final-state phase space is
\begin{equation}
\int d\Phi_2
=
\frac{1}{8\pi}\sqrt{1-\frac{4m_h^2}{s_{3\mathrm{body}}}}
=
\frac{1}{8\pi}\sqrt{1-\frac{4}{9N_E^2}},
\end{equation}
while
\begin{equation}
(2E_1)(2E_2)(2E_3)
=
(2N_E m_h)^3
=
8N_E^3 m_h^3.
\end{equation}
If we adopt the convention in which the symmetry factor $1/3!$ for the identical initial-state particles is included in the definition of the effective $3\to2$ cross section, then
\begin{equation}
\langle \sigma v^2\rangle_{3\to2}
\sim
\frac{1}{3!}\frac{1}{2!}
\frac{1}{(2E_1)(2E_2)(2E_3)}
\int d\Phi_2\,
\left|\mathcal M_{3\to2}^{(N_E)}\right|^2.
\end{equation}
This gives
\begin{equation}
\boxed{
\langle \sigma v^2\rangle_{3\to2}^{(N_E)}
\sim
\frac{1}{3!}\frac{1}{128\pi\,N_E^3 m_h^3}
\sqrt{1-\frac{4}{9N_E^2}}
\left|
10\,\frac{\lambda_3\lambda_4}{N_E^2 m_h^2}
+
15\,\frac{\lambda_3^3}{N_E^4 m_h^4}
\right|^2
.}
\label{eq:app_sigmav2_3to2_general}
\end{equation}

Substituting $\lambda_3=2\lambda\epsilon$ and $\lambda_4=6\kappa$, one may equivalently write

\begin{equation}
\boxed{
\langle \sigma v^2\rangle_{3\to2}^{(N_E)}
\sim
\frac{1}{3!}\frac{1}{128\pi}
\sqrt{1-\frac{4}{9N_E^2}}
\left(
\frac{14400\,(\lambda\epsilon)^2\kappa^2}{N_E^7 m_h^7}
+
\frac{28800\,(\lambda\epsilon)^4\kappa}{N_E^9 m_h^9}
+
\frac{14400\,(\lambda\epsilon)^6}{N_E^{11} m_h^{11}}
\right)
.}
\label{eq:app_sigmav2_3to2_expanded}
\end{equation}
\subsection{Prescription for the effective Mandelstam variable}
\label{app:effective_s_prescription}

In the main text, we do not solve the full momentum-dependent collision integrals. 
Instead, for each $2\to2$ process we evaluate the corresponding total cross section at an effective averaged Mandelstam invariant $\langle s\rangle$.

Starting from
\begin{equation}
s = m_1^2+m_2^2+2\left(E_1E_2-|\mathbf p_1||\mathbf p_2|\cos\theta\right),
\end{equation}
the exact statistical average is
\begin{equation}
\langle s\rangle
=
m_1^2+m_2^2
+
2\left(
\langle E_1E_2\rangle
-
\left\langle |\mathbf p_1||\mathbf p_2|\cos\theta \right\rangle
\right).
\label{eq:app_s_exact_average}
\end{equation}
Neglecting the anisotropic part of the angular distribution, we approximate
\begin{equation}
\left\langle |\mathbf p_1||\mathbf p_2|\cos\theta \right\rangle \simeq 0.
\end{equation}
If one further assumes weak correlations between the particle energies, then
\begin{equation}
\langle E_1E_2\rangle \simeq \langle E_1\rangle \langle E_2\rangle,
\end{equation}
and therefore
\begin{equation}
\langle s\rangle
\simeq
m_1^2+m_2^2+2\langle E_1\rangle\langle E_2\rangle .
\label{eq:app_s_average}
\end{equation}

For equal-mass incoming particles, $m_1=m_2=m$, this reduces to
\begin{equation}
\langle s\rangle \simeq 2m^2+2\langle E\rangle^2.
\end{equation}
We use $\langle s\rangle$ only as a characteristic center-of-mass energy scale to obtain an order-of-magnitude estimate of the cross section. It should be stressed that this treatment provides only an order-of-magnitude estimate. A more detailed determination of the post-collision momentum distribution lies beyond the scope of this work. 
For this reason, we adopt the above approximation and evaluate the cross section at a representative effective invariant \(\langle s\rangle\).

In practice, our algorithm is therefore:
\begin{enumerate}
    \item determine the characteristic average energy of the incoming particles. 
    Since there is currently no controlled first-principles result for the full distribution function of particles produced from bubble-wall collisions, we approximate their typical energy by the ratio of the total injected energy density to the total number density, namely
    \begin{equation}
        \langle E_i\rangle \equiv \frac{\rho_i}{n_i},
    \end{equation}
    for each species \(i\);
    
    \item construct the effective Mandelstam invariant \(\langle s\rangle\) using Eq.~\eqref{eq:app_s_average};
    
    \item evaluate the corresponding total cross section at \(s=\langle s\rangle\);
    
    \item form the effective interaction rate and compare it with the Hubble rate.
\end{enumerate}

For ordinary $2\to2$ processes, we use the estimate
\begin{equation}
\Gamma_{2\to2}\sim n\,\langle \sigma v\rangle,
\label{eq:app_rate_2to2}
\end{equation}
and compare
\begin{equation}
n\,\langle \sigma v\rangle
\quad \text{with} \quad H.
\end{equation}
By contrast, for the number-changing process $3h\to2h$, the rate scales as
\begin{equation}
\Gamma_{3\to2}\sim n_h^2\,\langle \sigma v^2\rangle,
\label{eq:app_rate_3to2}
\end{equation}
so in that case we compare
\begin{equation}
n_h^2\,\langle \sigma v^2\rangle
\quad \text{with} \quad H.
\end{equation}
This is the criterion we use to judge whether scalar self-interactions significantly deplete the $h$ abundance before it is converted into DM.

For the final parameter region selected in our analysis, the actual post-collision physical picture is therefore considerably simplified. Bubble collisions initially produce the three species $h$, $\psi$, and $\varphi_{\rm{inf}}$. After bubble collisions, the produced spectator particles \(h\) can undergo elastic self-scatterings,
\(hh\to hh\), which redistribute their momenta and may bring the \(h\) sector close to kinetic
equilibrium. Independently, the decay \(h\to\psi\bar\psi\) converts the spectator population into
dark matter, while all other competing inelastic processes remain negligible during the relevant epoch, in the sense that their reaction rates satisfy $\Gamma/H\ll 1$.

\section{Benchmark Points}
\label{markpoint}

\begin{table}[H]
\centering

\label{tab:benchmark_points}
\begin{tabular}{cccccc}
\toprule
$\mu \ [\mathrm{GeV}]$ & $\epsilon \ [\mathrm{GeV}]$ & $\kappa$ & $c_{\rm{inf}}$ & H$ \ [\mathrm{GeV}]$ & $m_{\psi}  \ [\mathrm{GeV}]$ \\
\midrule
6.164e+05 & 1.535e+06 & 1.800e+00 & 2.154e-12 & 1.000e+00 & 5.376e+04 \\
9.920e+06 & 2.007e+07 & 1.800e+00 & 2.783e-11 & 1.000e+00 & 4.379e+03 \\
1.531e+08 & 3.975e+08 & 1.800e+00 & 3.594e-10 & 1.389e+02 & 2.046e+01 \\
2.047e+08 & 5.313e+08 & 1.800e+00 & 3.594e-10 & 1.179e+01 & 1.102e+01 \\
2.513e+10 & 6.003e+10 & 1.800e+00 & 5.995e-08 & 2.276e+05 & 2.062e-01 \\
1.524e+12 & 2.960e+12 & 1.800e+00 & 7.743e-07 & 3.162e+07 & 2.102e+11 \\
\bottomrule
\end{tabular}

\caption{Selected benchmark points used in our numerical analysis. For each benchmark point, we list the corresponding values of $\mu$, $\epsilon$, $\kappa$, $c_{\rm{inf}}$, $H$, and $m_{\psi}$. All dimensionful quantities are given in GeV.}

\end{table}

\newpage

\bibliographystyle{JHEP}
\bibliography{biblio}

\end{document}